\documentclass[a4page, 10pt]{article}
\usepackage{amssymb,amsmath}
\usepackage{graphicx}
\numberwithin{equation}{section}

\begin{document}

\title{Wheeler's It from Bit Proposal in Loop Quantum Gravity}

\author{Jarmo M\"akel\"a\footnote{Vaasa University of Applied Sciences,
Wolffintie 30, 65200 Vaasa, Finland, email: jarmo.makela@vamk.fi}}

\maketitle

\begin{abstract}

As an attempt to realize Wheeler's "it-from-bit proposal" that physics should be reduced to simple yes-no questions we consider a model of loop quantum gravity, where the only allowed values of the quantum numbers $j_p$ at the punctures $p$ of the spin network on the spacelike two-surfaces of spacetime are $0$ and $\frac{1}{2}$. When $j_p=0$, the puncture is in the vacuum, and it does not contribute to the area of the two-surface, whereas when $j_p=\frac{1}{2}$, the puncture is in an excited state, and the allowed values of the associated quantum number $m_p$ are $-\frac{1}{2}$ and $+\frac{1}{2}$.  As a consequence, the spin network used as a model of spacetime is analogous to a system of particles with spin $\frac{1}{2}$, and every puncture carries exactly one bit of information. When applied to spacetimes with horizon our model enables us to find an explicit expression for the partition function of spacetime. Using this partition function we may, among other things, obtain the  Bekenstein-Hawking entropy law for black holes. When applied to cosmological models with horizon the partition function predicts a cosmic phase transition in the early Universe, where the cosmological constant went through a dramatic decrease and the matter of the Universe was created out of the vacuum.  

\end{abstract}

\bigskip

\section{Introduction}

John Archibald Wheeler (1911-2008) was one of the deepest thinkers of the twentieth century. Among other things, he was one of the first physicists to put forward serious attempts to quantize gravity: The spectacular second coming of general relativity in the 1960's was largely intiated by Wheeler. Through his PhD students he was influential in the development of  particle physics and the thermodynamics of black holes.

In his later years Wheeler's thoughts chrystallized to an idea of the concept of {\it information} as the basis of physics. This idea of his is best expressed in the famous essay based on his talk held in the year 1989 in the 3rd Symposium of Quantum Mechanics in Tokyo. \cite{yy} In the essay Wheeler explicitly rejects the concepts of space, time and the spacetime continuum at the microscopic level and, for the same reason, is very doubtful on the fundamental validity of the concept of wave function. Instead, he maintains that physics should be based on the elementary quantum pheneomena, where simple yes-no questions can be asked by the observer. In his own words, "every physical quantity, every it, derives its ultimate significance from bits, binary yes-no indications, a conclusion which we epitomize in the phrase {\it it from bit}".

The phrase "it from bit" captures the essential content of Wheeler's essay. When reading Wheeler's essay one cannot avoid the feeling that Wheeler wants to say more than he can express in words. The question is, whether Wheeler's overwhelming idea may be realized in a mathematically precise manner. Is there a model of the Universe, which both rejects space, time and spacetime continuum at the microscopic level, and reduces the information content of the Universe in bits?

One of the examples mentioned by Wheeler in his essay is black hole entropy. According to the Bekenstein-Hawking entropy law the entropy of the black hole is: \cite{kaa,koo,nee}
\begin{equation}
S = \frac{k_B}{4\ell_{Pl}^2}A,
\end{equation}
where
\begin{equation}
\ell_{Pl} := \sqrt{\frac{\hbar G}{c^3}} \approx 1.6\times 10^{-35}m
\end{equation}
is the Planck length, and $A$ is the event horizon area of the hole. If one partitions this area into domains, each of size $4\ell_{Pl}^2\ln(2)$, the number of the domains is
\begin{equation}
N = \frac{A}{4\ell_{Pl}^2\ln(2)},
\end{equation}
and we may express the black hole entropy as:
\begin{equation}
S = k_BN\ln(2).
\end{equation}
As a consequence, the number of the microscopic states of the hole is
\begin{equation}
\Omega = \exp\left(\frac{1}{k_B}S\right) = 2^N,
\end{equation}
which means that each domain carries exactly one bit of information hidden by the hole. The simple observation made by Wheeler suggests that 1) the event horizon area of the black hole is an integer times $4\ell_{Pl}^2\ln(2)$ and 2) each domain of the event horizon has two intrinsic quantum states. It is very tempting to take Wheeler's observation as the starting point of the realization of the it-from-bit proposal. 

There is a general agreement in the physics community that to explain the microscopic origin of black hole entropy a quantum theory of gravitation is needed. A serious candidate for the quantum theory of gravitation is  {\it loop quantum gravity}, which was put forward a bit more than 30 years ago. \cite{vii,kuu} The starting point of loop quantum gravity is an almost trivial observation that if we rotate the local frame of reference of an observer, the laws of physics in the rotated frame are exactly the same as they were in the original frame. As a consequence, one may re-formulate Einstein's general theory of relativity, at least formally, as a theory akin to  a gauge field theory with the rotation group $SO(3)$ as one of its local symmetry groups. Applying the standard rules of quantization to the resulting theory one obtains a quantum theory of gravitation. 

In loop quantum gravity spacetime is modelled by the so-called {\it spin network}. Hence loop quantum gravity abandons the spacetime continuum at the microscopic level, as requested by Wheeler. In very broad terms, spin network may be described as a graph lying on a spacelike hypersurface of spacetime such that every edge is asociated with a $(2j+1)$-dimensional representation of the group $SU(2)$, whose generators obey exactly the same Lie algebra as the generators of the group $SO(3)$ $(j= 0, \frac{1}{2}, 1, \frac{3}{2},\dots)$. \cite{seite} The most important single prediction of loop quantum gravity is that the area of any spacelike two-surface has a discrete spectrum. More precisely, it turns out that the possible area eigenvalues of the given spacelike two-surface are, in the SI units, of the form:
\begin{equation}
A = 8\pi\gamma\ell_{Pl}^2\sum_p\sqrt{j_p(j_p+1)},
\end{equation}
where $\gamma$ is a pure number, which is known as the Immirzi parameter. In Eq. (1.6) we have summed over the {\it punctures} $p$ of the spin network on the two-surface, or the edges, which intersect the two-surface. The quantum numbers $j_p$ determine the dimension of the representation of the group $SU(2)$ associated with the puncture $p$: For given $j_p$ there exists the quantum number $m_p$ such that
\begin{equation}
m_p \in \lbrace-j_p, -j_p +1, \dots, j_p-1, j_p\rbrace,
\end{equation}
and hence for given $j_p$ there are $2j_p +1$ possible values of $m_p$. 

In this paper we consider a possiblity that the only allowed values taken by the quantum numbers $j_p$ at the punctures $p$ of the spin network are $0$ and $\frac{1}{2}$. When $j_p = 0$, the puncture does not contribute to the area of the spacelike two-surface, and we say that the puncture is in the {\it vacuum}, whereas if $j_p = \frac{1}{2}$, the puncture is in an excited state. This idea of ours is in a marked contrast with the standard loop quantum  gravity, where $j_p$ may be any non-negative integer or half-integer. An advantage of our model is that it allows us to make a connection with Wheeler's it-from-bit proposal:  When $j_p=\frac{1}{2}$ and the puncture is in an excited state, then $m_p$ is either $-\frac{1}{2}$ or $+\frac{1}{2}$, and hence there are {\it just two} possible values taken by $m_p$. In this sense the punctures in the excited state are somewhat akin to spin $\frac{1}{2}$ particles, which may have their spins either up or down, and every puncture carries exactly one bit of information, as requested by Wheeler. Moreover, if we choose
\begin{equation}
\gamma = \frac{\ln(2)}{\sqrt{3}\pi}
\end{equation}
in Eq. (1.6), and denote by $N$ the number of the punctures in the excited states at the black hole event horizon, the horizon area is
\begin{equation}
A = 4N\ell_{Pl}^2\ln(2),
\end{equation}
and Eq. (1.3) is recovered.

Even though the basic idea of this paper is very simple, its systematic development requires very great care. For instance, to consider the entropic properties of spacetime one should be able to write an explicit expression for its partition function. To be able to write the partition function, in turn, one should determine the energy of spacetime, and because the concept of energy is observer-dependent, the very first thing to do is to specify the observer. Surprisingly, we shall see in this paper that it is indeed possible to carry out this program explicitly for any stationary spacetime with closed and compact horizon. Those spacetimes include, among other things, the Kerr-Newman spacetime, which is the most general stationary spacetime involving  a black hole, together with the de Sitter spacetime. The most important specific result of our analysis is that at a certain characteristic temperature $T_C$ spacetime performs a {\it phase transition}, where the punctures of the spin network at the horizon jump from the vacuum, where $j_p=0$ to the excited states, where $j_p=\frac{1}{2}$. For an observer with constant proper acceleration $a$ cölose to the horizon the characteristic temperature $T_C$ equals with the Unruh temperature measured by the observer. The phase transition may be used to explain the thermal properties of black holes, and it has important implications regarding the properties of the cosmological constant and dark energy. Finally, the phase transition tells how the information carried by the matter in the observable Universe may be reduced in bits, thus realizing at least part of Wheeler's it-from-bit proposal. 

Unless otherwise stated, we shall always use the natural units, where $\hbar = c = G = k_B = 4\pi\epsilon_0 = 1$.   

\section{Observers}

As it is well known, the clue which led Albert Einstein to his general theory of relativity was his realization in the year 1907 that, according to his own words, "if a man falls freely, he would not feel his own weight". \cite{kasi} One of the consequences of this idea, which Einstein later developed into his equivalence principle, is that whether an observer feels effects generally attributed to gravity depends on his state of motion or, to be more precise, on his world line in spacetime. In particular, the effects observed in  local measurements in a uniformly accelerated frame of reference are indistinguishable from the effects observed in the uniform gravitational field. In this sense observers with the same acceleration may be considered equivalent. 

It appears to the author that Einstein's principle of equivalence has not received sufficient attention in the attempts to quantize gravity. After all, in quantum mechanics there is not, strictly speaking, any absolute quantum state of the system, but we may only talk about the quantum state of the system with respect to this or that observer. Such a view has been advocated, in particular, by Rovelli in his relational interpretation of quantum mechanics. \cite{kasiaa} When an observer performs a measurement on a system, the quantum state of the system collapses, from the point of view of our specific observer, to one of the eigenstates of the observable measured by the observer. 

The equivalence of observers with the same acceleration in general relativity suggests that we cannot talk meaningfully about the quantum state of any system without a reference to the acceleration or, more precisely, to the {\it proper acceleration} of the observer. Indeed, even when treated non-relativistically, the quantum mechanics of the free particle in an accelerated frame is completely different from its quantum mechanics in an inertial frame. The quantum theory of fields, in turn, has taught us that an observer with proper acceleration $a$ detects a flow of particles with {\it Unruh temperature} \cite{ysi}
\begin{equation}
T_U := \frac{a}{2\pi}
\end{equation}
even when the inertial observers detect no particles at all. In other words, the concept of particle is observer-dependent, and the properties of the particles depend on the proper acceleration of the observer.

Of particular importance is the specification of the proper acceleration of the observer, when we attempt to determine the quantum state of the gravitational field which, in some sense, may be viewed as the spacetime itself. In curved spacetime the proper acceleration vector field of the congruence of the world lines of observers is, by definition,
\begin{equation}
a^\mu := u^\alpha u^\mu_{;\alpha},
\end{equation}
where $u^\mu$ is the future pointing unit tangent vector field of the congruence. As usual, the semicolon means covariant differentiation. An observer with proper acceleration $a^\mu$ measures the gravitational acceleration 
\begin{equation}
a := \sqrt{a_\mu a^\mu}
\end{equation}
for particles in a free fall. For instance, in the Schwarzschild spacetime, where the line element is:
\begin{equation}
ds^2 = -\left(1 - \frac{2M}{r}\right)\,dt^2 + \frac{dr^2}{1 - \frac{2M}{r}} + r^2\,d\theta^2 + r^2\sin^2(\theta)\,d\phi^2,
\end{equation}
the only non-zero component of the future pointng unit tangent vector field of the congruence of the world lines of the observers with constant coordinates $r$, $\theta$ and $\phi$ is:
\begin{equation}
u^t = \left(1 - \frac{2M}{r}\right)^{-1/2}.
\end{equation}
The only non-zero component of the proper acceleration vector field $a^\mu$ is
\begin{equation}
a^r = u^t u^r_{;t} = u^t\Gamma^r_{tt}u^t = \frac{M}{r^2},
\end{equation}
and therefore an observer with constant $r$, $\theta$ and $\phi$ measures the gravitational acceleration
\begin{equation}
a = \sqrt{a_r a^r} = \left(1 - \frac{2M}{r}\right)^{-1/2}\frac{M}{r^2}
\end{equation}
for particles in the radial free fall. If one accepts the view that the determination of the quantum state of requires the specification of the proper acceleration of the observer, one is led to the conclusion that the proper acceleration $a$ of the observer should be kept as a constant during the measurement process. Keeping the proper acceleration of the observer as a constant during the measurement we may make sure that the observed changes in the the state really reflect the changes in the physical quantum  state of the spacetime, rather than the changes in the frame of reference of the observer.

To illustrate this idea consider, as an example, an observer originally at rest with respect to the Schwarzschild coordinates $r$, $\theta$ and $\phi$ in spacetime involving  a Schwarzschild black hole with Schwarzschild mass $M$. If the mass $M$ of the hole takes an infinitesimal change $dM$ (for instance, the hole may aquire more matter, or it may evaporate by means of the Hawking radiation) we must change the radial coordinate $r$ of the observer by $dr$ to keep the proper acceleration $a$ of the observer as a constant. Between the changes $dM$ and $dr$ there is the relationship:
\begin{equation}
da = \frac{\partial a}{\partial M}\,dM + \frac{\partial a}{\partial r}\,dr = 0,
\end {equation}
and Eq. (2.7) implies:
\begin{equation}
\frac{1}{r}\,dM - \frac{M}{r^2}\,dr = \left(1 - \frac{2M}{r}\right)\left(2\frac{dr}{r} - \frac{dM}{M}\right).
\end{equation}
Just outside of the event horizon, where $r=2M$, the right hand side of Eq. (2.9) will effectively vanish, and we have, to a good approximation:
\begin{equation}
dr = 2\,dM.
\end{equation}
This result means that an observer originally close to the horizon will stay close to the horizon, when the mass $M$ of the hole is changed.

\section{Energy}

Information theory and thermodynamics are closely related. For instance, the thermodynamical entropy of any system is proportional to its Shannon entropy \cite{kymppi} which, in turn, tells the number of bits needed to specify the state of the system. Unfortunately, the thermodynamical considerations of any system are based on its {\it energy} which, as it was observed by Wheeler in Ref. \cite{yytoo}, is very problematic in general relativity. However, for static spacetimes it is possible to define the so-called {\it Brown-York energy} \cite{kaatoo}
\begin{equation}
E_ {BY} := -\frac{1}{8\pi}\oint_{\mathcal{S}}(k-k_0)\,d\mathcal{A}.
\end{equation}
In this equation we have integrated over a closed two-surface $\mathcal{S}$ embedded into a spacelike hypersurface of spacetime, where the time coordinate $t = constant$. $d\mathcal{A}$ is the area element on this two-surface, and  $k$ is the trace of of the exterior curvature tensor induced on the two-surface. $k_0$ is the trace of the exterior curvature tensor, when the two-surface has been embedded into flat spacetime. 

The concept of Brown-York energy is very natural, and it stems from the Hamilton-Jacobi formulation of general relativity. One may view the Brown-York energy as the energy of the gravitational field inside of the closed two-surface $\mathcal{S}$. It is particularly interesting to consider the properties of the Brown-York energy, when the closed two-surface is in the immediate vicinity of a {\it horizon} of spacetime. The horizon of spacetime may be either the event horizon of a black hole, or the cosmological horizon of an expanding universe. Very interesting general results for the properties of the Brown-York energy of spacetime may be obtained, if we consider the effects of the {\it change} of the parameters of the spacetime on its Brow-York energy.

As it is well known, the most general stationary black hole solution to the combined Einstein-Maxwell equations in vacuum is the {\it Kerr-Newman solution}, which describes spacetime involving the so-called Kerr-Newman black hole. The Kerr-Newman black hole is completely determined by just {\it three parameters}, which are the mass $M$, angular momentum $J$, and the electric charge $Q$ of the hole. Even though the Kerr-Newman spacetime is not exactly static, but just stationary, we shall see in Appendix A of this paper that it is possible to pick up in the Kerr-Newman spacetime a system of coordinates, where the geometry of spacetime is effectively static just outside of the event horizon of the Kerr-Newman black hole. This enables us to define the Brown-York energy for the Kerr-Newman black holes.

When considering the effect of the change of the parameters $M$, $J$ and $Q$ on the Brown-York energy of the Kerr-Newman spacetime we employ the important conclusion drawn in  Section 2 that the proper acceleration $a$ of the observer should be kept as a constant during the measurement process. We shall also assume that the observer lies very close to the event horizon of the Kerr-Newman black hole. For the sake of simplicity we shall call the closed spacelike two-surface, where the proper acceleration  $a$ is a constant everywhere, no matter what may happen to the parameters $M$, $J$ and $Q$, just outside of the event horizon of the hole, as the {\it stretched horizon} of the Kerr-Newman black hole. When the parameters $M$, $J$ and $Q$ are changed, the stretched horizon will also change, but in such a way that the proper acceleration $a$ stays unchanged everywhere on the stretched horizon. One of the reasons for considering the Brown-York energy of the Kerr-Newman black hole from the point of view of an observer on the  stretched horizon $a = constant$ is that in this way we may ignore, among other things, the backscattering effects of the Hawking radiation from the spacetime geometry. 

In Appendix A of this paper we have proved a very important general result that when the parameters $M$, $J$ and $Q$ undergo infinitesimal changes $dM$, $dJ$ and $dQ$, then between the resulting change $dA$ in the stretched horizon area, and the corresponding change $dE_{BY}$ in the Brown-York energy there is the relationship:
\begin{equation}
dE_{BY} = \frac{a}{8\pi}\,dA.
\end{equation}
In Ref. \cite{kootoo} this result was proved for the Schwarzschild black hole, in Ref. \cite{neetoo} for the Reissner-Nordstr\"om black hole, and in Ref. \cite{viitoo} for the general, spherically symmetric spacetime with horizon. Since the Kerr-Newman black hole is the most general stationary black hole in Einstein's general theory of relativity, we have thus managed to prove that Eq. (3.2) is a generic result, which holds for {\it all} black holes. We have shown in Appendix A that a stretched horizon, where $a=constant$ originally close to the event horizon will stay close to the event horizon, no matter what may happen to the parameters $M$, $J$ and $Q$. Hence we may identify, in effect, the area $A$ of the stretched horizon with the event horizon area of the hole. 

During the formation of the black hole by means of the gravitational collapse matter will flow through the stretched horizon, and its area $A$ is increased from zero to its final value. The proper acceleration $a$ stays unchaged during the process, and therefore we may view the quantity
\begin{equation}
E = \frac{a}{8\pi}A
\end{equation}
 as the energy of the Kerr-Newman black hole from the point of view of an observer at rest on the stretched horizon, where $a=constant$.
  
As one may observe, the expression for the energy of the general black hole in Eq. (3.3) is remarkably simple. Its simplicity arises from our decision, motivated by the equivalence principle, to consider the black hole from the point of view of an observer with constant proper acceleration $a$. Actually, Eq. (3.3) is a classic result, which was first obtained by Frodden, Gosh and Perez in Ref. \cite{viitooaa}. The novelty of our approach is the derivation of Eq. (3.3) from the notion of the Brown-York energy.

 A somewhat similar result may be obtained for the energy of the de Sitter spacetime as well. By definition, the de Sitter spacetime is an empty, spherically symmetric spacetime with positive cosmological constant $\Lambda$. In the static coordinates the line element of the de Sitter spacetime takes the form: \cite{kuutoo}
\begin{equation}
ds^2 = -\left(1 - \frac{\Lambda}{3}r^2\right)\,dt^2 + \frac{dr^2}{1 - \frac{\Lambda}{3}r^2} + r^2\,d\theta^2 + r^2\sin^2(\theta)\,d\phi^2.
\end{equation}
 The de Sitter spacetime has the {\it cosmological horizon}, where
\begin{equation}
r = r_C := \sqrt{\frac{3}{\Lambda}}.
\end{equation}
When $r<r_C$, spacetime is static, and one may write the Brown-York energy for the gravitational field inside of the given closed, spacelike two-surface of spacetime. As an analog of the stretched horizon of the black hole one may define the {\it shrinked horizon} of the de Sitter spacetime as a closed spacelike two-surface, just inside of the cosmological horizon, where the proper acceleration $a=constant$. It was shown in Ref. \cite{seetoo} that if the cosmological constant $\Lambda$ is allowed to change, then between the resulting change $dA$ in the shrinked horizon area $A$ and the corresponding change $dE_{BY}$ in the Brown-York energy $E_{BY}$ there is the relationship:
\begin{equation}
dE_{BY} = -\frac{a}{8\pi}\,dA.
\end{equation}
The minus sign on the right hand side of this equation indicates that energy is flown {\it outside} of the shrinked horizon, when its area is increased. Actually, this is something one might have expected, since the cosmological constant $\Lambda$ may be understood as a quantity, which is proportional to the energy density of the vacuum. Eq. (3.5) implies that when the area of the shrinked horizon increases, the cosmological constant must {\it decrease}, and hence the vacuum energy inside of the shrinked horizon must decrease during the increase of the shrinked horizon. Nevetheless, if we still understand the energy of spacetime, as for black holes, as the gravitational energy flown accross the horizon during its formation, we may take the quantity
\begin{equation}
E = \frac{a}{8\pi}A
\end{equation}
as the energy of the de Sitter spacetime. This energy is identical to the black hole energy introduced in Eq. (3.3), and hence it appears that we have managed to obtain for the gravitational energy an expression, which holds not only for all possible black holes, but for spacetimes equipped with the cosmological horizon as well. Eq. (3.3) is therefore very important, and it provides a starting point for our forthcoming discussion.

\section{Energy Operator}

In his essay Wheeler explicitly rejects the idea of spacetime as a continuum. \cite{yy} Actually, there does exist a quantum theory of gravity, which really relies on discrete, rather than continuous structures. That theory is known as {\it loop quantum gravity}. \cite{vii,kuu} In loop quantum gravity spacetime is described by the so-called {\it spin network}. In broad terms, spin network is a graph lying on the spacelike hypersurface of spacetime, where the time coordinate $t=constant$. \cite{seite} With each edge of the graph one associates an irreducible representation of the group $SU(2)$. When passing from classical general relativity to loop quantum gravity one replaces certain classical quantities by the corresponding quantum-mechanical operators. The area $A$ of the given spacelike two-surface of spacetime, for instance, is replaced by the corresponding {\it area operator}
\begin{equation}
\hat{A} := 8\pi\gamma\ell_{Pl}^2\sum_p\sqrt{\hat{J}^s(p)\hat{J}_s(p)},
\end{equation}
where $\ell_{Pl} :=\sqrt{\frac{\hbar G}{c^3}}$ is the Planck length, and the pure number $\gamma$ is known as the {\it Immirzi parameter}. We have summed over the punctures $p$ of the spin network on the two-surface, and the Hermitean operators $\hat{J}_s(p)$ $(s=1,2,3)$ are the generators of the group $SU(2)$ at the puncture $p$. As it is well known from the elementary courses of quantum mechanics, the operator
\begin{equation}
\hat{\vec{J}}^2(p) := \hat{J}^s(p)\hat{J}_s(p),
\end{equation}
and one of the operators $\hat{J}_s(p)$, say $\hat{J}_3(p)$, have common eigenstates $\vert j_p m_p\rangle$ such that
\begin{subequations}
\begin{eqnarray}
\hat{\vec{J}}^2(p)\vert j_p m_p\rangle &=& j_p(j_p+1)\vert j_p m_p\rangle,\\
\hat{J}_3(p) &=& m_p\vert j_p m_p\rangle,
\end{eqnarray}
\end{subequations}
where $j_p\in\lbrace 0,\frac{1}{2},1,\frac{3}{2},2,\dots\rbrace$, and $m_p \in\lbrace -j_p, -j_p+1,\dots, j_p -1,j_p\rbrace$ for fixed $j_p$. This means that for fixed $j_p$ we associate a $(2j_p +1)$-dimensional representation space of the group $SU(2)$ with the puncture $p$. 

Eqs. (3.3) and (4.1) imply that the Hamiltonian-, or the energy operator of spacetime takes, from the point of view of an observer either on a stretched or a shrinked horizon of spacetime, where the proper acceleration $a=constant$, the form:
\begin{equation}
\hat{H} = \gamma a\sum_p\sqrt{\hat{J}^s(p)\hat{J}_s(p)}.
\end{equation}
One might feel worried because of the presence of the proper acceleration $a$ on the right hand side of Eq. (4.4). After all, the Hamiltonian operator $\hat{H}$ is supposed to describe the microscopic, Planck-size properties of spacetime, whereas the proper acceleration $a$ clearly is a macroscopic, classical quantity. Such a worry, however, is unnecessary. In the standard Copenhagen interpretation of quantum mechanics the observers are always assumed to use macroscopic, classical instruments for the measurements. The interaction between the system and the instrument makes the state vector of the system to collapse to one of the eigenstates of the measured quantity. As we concluded in Section 2, in quantum gravity we must always refer to the proper acceleration of the observer. Hence the presence of the proper acceleration $a$ on the right hand side of Eq. (4.4) is actually exactly what one expects.

Assuming that there are $N$ punctures of the spin network on the stretched or shrinked horizon of spacetime one immediately observes that the state
\begin{equation}
\vert\psi\rangle := \vert j_1 m_1\rangle\otimes\vert j_2m_2\rangle\otimes\cdots\otimes\vert j_Nm_N\rangle
\end{equation}
is an energy eigenstate of spacetime. The corresponding energy eigenvalue is, as it follows from Eq. (4.3a):
\begin{equation}
E = \gamma a\sum_{p=1}^N\sqrt{j_p(j_p+1)},
\end{equation}
where $j_p\in\lbrace 0,\frac{1}{2},1,\frac{3}{2}\dots\rbrace$ for all $p=1,2,\dots,N$. If $j_p=0$, we say that the puncture $p$ is in {\it vacuum}; otherwise we say that it is in an {\it excited state}. 

The question is now: What are the possible values of the quantum numbers $j_p$? Whe facing with this question we are in a somewhat similar situation as we are, when attempting to predict the electron spin by means of the general postulates of quantum mechanics: The general postulates of quantum  mechanics merely imply that the electron spin is an integer or a half-integer, and for the exact determination of the electron spin observational data is needed. In loop quantum gravity we do not currently have any direct observational data for the determination of the quantum numbers $j_p$ (except possibly, as we shall see in Secs. 8 and 9, the accelerating expansion of the Universe), but we have nevertheless the well-known results obtained from the quantum field theory in curved spacetime for the black hole radiation and the entropy of the de Sitter spacetime. Whatever choice we may make for the possible values of $j_p$, the consequences of this choice should be consistent with those results. Another clue to the possible values of $j_p$ is provided by the general tendency of nature to the greatest possible simplicity. The electron spin, for instance, is just $\frac{1}{2}$, and no elementary particles with spin higher than $1$ have been observed in nature. The simplest possible non-trivial choice is to take 0 and $\frac{1 }{2}$ as the only allowed values of the quantum number $j_p$. When $j_p=0$, the puncture $p$ is in vacuum, and it does not contribute to the area of the horizon, whereas when $j_p=\frac{1}{2}$, the puncture $p$ will contribute an elementary area
\begin{equation}
A_0 := 4\sqrt{3}\pi\gamma\ell_{Pl}^2
\end{equation}
to the horizon. One of the attractive features of this simplest possible choice of $j_p$ is that the area of the horizon is now of the form:
\begin{equation}
A_n = n\cdot4\sqrt{3}\pi\gamma\ell_{Pl}^2,
\end{equation}
where $n=0,1,2,\dots$ is the number of the punctures in the excited states. Among other things, Eq. (4.8) implies that the event horizon of the black hole has an {\it equal spacing} in its area spectrum. An equally spaced area spectrum for the event horizon of the black hole was proposed by Bekenstein already in the year 1974. \cite{kasitoo} Bekenstein's original proposal was revived by Bekenstein and Mukhanov in 1995.  \cite{ysitoo} Since then, equally spaced area spectra for the black holes have been proposed by several authors on various grounds. \cite{kakskyt}

What makes the choice $j_p=\frac{1}{2}$ for the only possible non-vacuum value of $j_p$ really attractive, however, is the fact that when $j_p=\frac{1}{2}$, the quantum number $m_p$ may take {\it exactly two} possible values. Those values are $+\frac{1}{2}$ and $-\frac{1 }{2}$. Each puncture of the spin network on the stretched or shrinked horizon of spacetime may thus have just two possible non-vacuum quantum states. In this sense we have managed to reduce the information carried by the quantum states of the horizon, and hence by the states of the spacetime itself, in bits, as requested by Wheeler in his essay. The punctures of the spin network on the stretched or the shrinked horizon of spacetime are somewhat analogous to electrons: The quantum theory of fields has taught us that electrons may now and then be created out of the vacuum, and each of those electrons has spin up or down with respect to the given magnetic field, whereas in  our model  the punctures may now and then jump from the vacuum to an  excited state, and each puncture in an excited state has exactly two linearly independent quantum  states. Because of that we may view the gravitational field as a system of spin $\frac{1}{2}$ objects. The "spin" of an excited state is always either "up" or "down", and hence each puncture carries exactly one bit of information. We shall see in Sections 7 and 8 how our choice to restrict the possible values of $j_p$ to $0$  and $\frac{1}{2}$ predicts for black holes and the de Sitter spacetime thermal properties, which are consistent with the standard results obtaned from the quantum field theory in curved spacetime, whereas if arbitrary values of $j_p$ were allowed, the predictions of our model would be slightly different from those results. 

Before closing our discussion about the energy operator and the energy eigenstates of spacetime a few words about the statistics of spacetime are necessary. In quantum mechanics identical particles with integer spin obey the Bose-Einstein statistics, which means that interchange of any two particles keeps the overall quantum state of the system unchanged, whereas identical particles with half-integer spin obey the Fermi-Dirac statistics which, in turn, means that interchange of two particles will change the overall sign of the state vector. The reason for this fundamental property of identical particles lies deep down in the quantum theory of fields, and ultimately it may be reduced to the symmetries of flat spacetime. \cite{kaayy} In quantum gravity, however, we consider spacetime at the Planck length scale. At the Planck length scale spacetime is presumably totally different from flat spacetime, and no symmetries akin to the symmetries of the flat spacetime may be expected to exist. As a consequence, there is no reason why the punctures of the spin network on the stretched or shrinked horizon should obey statistics akin to the Bose-Einstein, or the Fermi-Dirac statistics, either. Rather, the contrary is the case, and one expects that every time, when we interchange two punctures at different quantum states, the overall quantum state of the spacetime will also change. Because of that we shall always write the energy eigenstates of spacetime as in Eq. (4.5) such that $j_p$ is always either $0$ or $\frac{1}{2}$ for all $p = 1,2,\dots,N$. Indeed, if we interchange two punctures at different quantum states, the overall energy eigenstate will also change. We shall see in the next section that this choice of ours will simplify considerably the calculation of the partition function of spacetime.  

\section{The Partition Function}

The partition function of any system is, by definition,
\begin{equation}
Z(\beta) := \sum_n\exp(-\beta E_n).
\end{equation}
In this definition we have summed over the energy eigenstates $n$ of the system. The quantities $E_n$ are the energy eigenvalues, and $\beta$ is the temperature parameter. In Section 4 we noticed an analogy of our system of the punctures of the spin network with the system of electrons. Whenever we calculate a partition function for a system of electrons, we take into account the electrons in the excited states only, {\it i. e} the electrons, which contribute to the total energy of the system. Likewise, when we obtaining an expression for the partition function of our system of punctures, we take into account the punctures in the non-vacuum states only. Eq. (4.6) implies that whenever a puncture is in a non-vacuum state, it contributes an energy 
\begin{equation}
E_0 = \frac{\sqrt{3}}{2}\gamma a
\end{equation}
to the spacetime. Hence it follows that the possible energy eigenvalues of spacetime from the point of view of our observer are of the form:
\begin{equation}
E_n = n\frac{\sqrt{3}}{2}\gamma a,
\end{equation}
where $n$ is the number of the punctures in the non-vacuum states. If the total number of the punctures on the stretched or shrinked horizon is $N$, the partition function $Z(\beta)$ in Eq. (5.1) takes the form:
\begin{equation}
Z(\beta) = \sum_{n=1}^N 2^n\exp\left(-n\beta\frac{\sqrt{3}}{2}\gamma a\right),
\end{equation}
where the factor $2^n$ comes from the fact that each puncture $p$ has two possible non-vacuum states, both with the same energy, where $m_p$ is either $+\frac{1}{2}$ or $-\frac{1}{2}$. 

It is now very easy to obtain an explicit expression for the partition function of spacetime from the point of view of our observer. To begin with, we define a new variable
\begin{equation}
z := 2^{\beta T_C - 1},
\end{equation}
where we shall call the temperature
\begin{equation}
T_C := \frac{\sqrt{3}\gamma a}{2\ln(2)}
\end{equation}
as the {\it characteristic temperature} of spacetime. In the low-temperature limit, where the temperature parameter $\beta$ tends to infinity, the variable $z$ tends to infinity as well, whereas in the high-temperature limit, where $\beta$ tends to zero, the variable $z$ tends to $\frac{1}{2}$. It is more convenient to write the partition function as a function of $z$, rather than as a function of $\beta$. We find:
\begin{equation}
Z(z) = \frac{1}{z} + \frac{1}{z^2} + \cdots + \frac{1}{z^N}.
\end{equation}
This is a simple geometrical sum written for $\frac{1}{z}$. We get:
\begin{equation}
Z(z) = \frac{1}{z-1}\left(1 - \frac{1}{z^N}\right).
\end{equation}
All thermodynamical properties of spacetime follow from this partition function. Eq. (5.8) is valid, whenever $z\ne 1$. If $z=1$, which means that the temperature $T$ equals with the characteristic temperature $T_C$, Eq. (5.7) implies:
\begin{equation}
Z(z) = N.
\end{equation}
We shall see later that the number $N$ of the punctures of the spin network on the stretched or shrinked horizon of spacetime, which is assumed to be fixed and very large, plays an important role in our discussion. 

   It should be noted that a somewhat similar calculation of the partition function has been carried out in Refs. \cite{kaayya} and \cite{kaayyb}. However, there are fundamental differences between the treatments in this paper, and the one used in Refs. \cite{kaayya} and \cite{kaayyb}: In Refs. \cite{kaayya} and \cite{kaayyb} the quantum numbers $j_p$ are allowed to take any integer or half-integer values, whereas in our paper the only allowed values of $j_p$ are $0$ and $\frac{1}{2}$. Even more important difference is that in Refs. \cite{kaayya} and \cite{kaayyb} the punctures are considerd indistinguishable, whereas in this paper that assumption is rejected for the reasons explained at the end of Section 4. As a consequence, the statistics of the punctures, and hence the partition function, is completely different from ours.

\section{Phase Transition}

The energy of any system at the given temperature $T = \frac{1}{\beta}$ is:
\begin{equation}
E(\beta) =  - \frac{\partial}{\partial \beta}\ln Z(\beta).
\end{equation} 
Eqs. (5.5) and (5.8) imply that we may write the energy $E$ as a function of the variable $z$ as:
\begin{equation}
E(z) =  \left(\frac{z}{z-1} + \frac{N}{1 - z^N}\right)T_c\ln(2),
\end{equation}
whenever $z\ne 1$. We have shown in Appendix B that
\begin{equation}
E(z) = \frac{1}{2}(N+1)T_C\ln(2),
\end{equation}
when $z=1$.

Consider now Eq. (6.2) in details. In the low-temperature limit the temperature parameter $\beta$, and hence $z$, will tend to infinity. Since
\begin{equation}
\lim_{z\rightarrow\infty}\left(\frac{z}{z-1} + \frac{N}{z^N - 1}\right) = 1,
\end{equation}
we find that in the low-temperature limit
\begin{equation}
E(z) = T_C\ln(2) = \frac{\sqrt{3}}{2}\gamma a,
\end{equation}
where we have used Eq. (5.6). Comparing Eqs. (5.2) and (6.5) we observe that in the low-temperature limit the average energy of spacetime equals with the energy contributed by a single puncture to the energy of spacetime, when that puncture is in the excited state, where $j_p = \frac{1}{2}$. This means that when the temperature of spacetime is very low from the point of view of our observer, just one of the punctures of the spin network is in the excited state, whereas all of the other punctures are in the vacuum, where $j_p=0$. In this limit the area of the stretched or shrinked horizon is just the elementary area
\begin{equation}
A_0 = 4\sqrt{3}\pi\gamma\ell_{Pl}^2
\end{equation}
of Eq. (4.7). In other words, the horizon willl effectively vanish. 

To consider the properties of spacetime outside of the low-temperature limit let us differentiate the energy  $E $ of Eq. (6.2) with respect to the absolute temperature $T$. We get:
\begin{equation}
\frac{dE}{dT} = \frac{dE}{dz}\frac{dz}{dT} = \left[\frac{z}{(z-1)^2} - \frac{N^2z^N}{(1-z^N)^2}\right]\frac{[\ln(2z)]^2}{\ln(2)}.
\end{equation}
It should be remembered that $N$ is assumed to be very large, indeed. For a typical astrophysical black hole, for instance, $N$ is around $10^{80}$, or so. In the large $N$ limit $z^N$ tends very rapidly to zero, whenever $z<1$, which means that the second term inside of the brackets on the right hand side of Eq. (6.7) will vanish. The second term will also vanish, when $z>1$, because in that case the second term is, in the leading order approximation for large $N$, $-N^2z^{-N}$, which definitely vanishes in the large $N$ limit. Hence we may write, in effect, 
\begin{equation}
\frac{dE}{dT} = \frac{z}{(z-1)^2}\frac{[\ln(2z)]^2}{\ln(2)},
\end{equation}
whenever $z\ne 1$. In the high-temperature limit, where $z$ tends to $\frac{1}{2}$, $\frac{dE}{dT}$ will vanish altogether. This means that at high temperatures the energy $E$ is essentially a constant function of the absolute temperature $T$. 

Something strange, however, will happen, when $z=1$, which means, through Eq. (5.5), that $T=T_C$. In Appendix B we have shown that
\begin{equation}
\frac{dE}{dT}\bigg\vert_{T=T_C} = \frac{1}{12}N^2[\ln(2)]^2 + O(N),
\end{equation}
where $O(N)$ denotes the terms proportional to the first or lower powers of $N$. Hence we observe that for large $N$ $\frac{dE}{dT}$ becomes enormous, when the temperature $T$ of spacetime from the point of view of our observer equals with its characteristic temperature $T_C$. This indicates a {\it phase transition} at the characteristic temperature $T_C$. During the phase transition there is an enormous jump in the energy $E $ of spacetime. Indeed, if we look at Eq. (6.2), we observe that whenever $z>1$, which means that $T<T_C$, we have, in effect:
\begin{equation}
E(z) = \frac{z}{z-1}T_C\ln(2).
\end{equation}
This follows from the fact that when $z>1$, $\frac{N}{z^N}$ tends to zero for large $N$, and therefore the second term inside of the brackets on the right hand side of Eq. (6.2) will vanish. However, when $z<1$, which means that $T>T_C$, $z^N$ tends rapidly to zero for large $N$. As a consequence, the second term inside of the brackets will vastly dominate over the first term, and we may write, in effect,
\begin{equation}
E = NT_C\ln(2).
\end{equation}
Hence the energy is around $T_C\ln(2)$ before the phase transition at the temperature $T_C$, and around $NT_C\ln(2)$ after the phase transition. For large $N$ this jump in the energy during the phase transition is enormous. It is interesting to observe that when $T>T_C$, the energy $E$ is, in effect, a {\it constant function} of the temperature $T$. This is something one might have expected since, as we saw in Eq. (6.8), $\frac{dE}{dT}$ tends to zero for large $T$. 

What will happen to the punctures during the phase transition? As we saw above, in the low-temperature limit just one of the punctures $p$ is in the excited state, where $j_p = \frac{1}{2}$, whereas the other punctures are in the vacuum, where $j_p = 0$. Eq. (6.10) implies that increase in the temperature $T$ does not produce an appreciable change in the energy $E$, when $T<T_C$. In other words, the punctures are effectively in the vacuum, when $T<T_C$. After the phase transition at the temperature $T=T_C$ has been completed, however, the energy is given by Eq. (6.11), and using Eq. (5.6) we find:
\begin{equation}
E =N\frac{\sqrt{3}}{2}a\gamma.
\end{equation}
Comparing Eqs. (5.2) and (6.12) we observe that on the right hand side of Eq. (6.12) we have exactly $N$ times the energy contributed by a single puncture in the excited state, where $j_p=\frac{1}{2}$ to the energy of the spacetime. This means that after the phase transition has been completed, all $N$ punctures of the spin network on the stretched or shrinked horizon have jumped, in effect, from the vacuum to the excited state.

Since the punctures are effectively in the vacuum, when $T<T_C$, the horizon is a Planck-size object, and in practice there is not a horizon at all. In this sense we may regard the characteristic temperature $T_C$ as the lowest possible temperature of the spacetime from the point of view of our observer. At the phase transition temperature $T_C$ the horizon turns from microscopic to macroscopic. On the other hand, however, we may view $T_C$ as the highest possible temperature of spacetime as well. This conclusion may be drawn from the fact that when $T>T_C$, the energy of spacetime is essentially a constant function of the temperature $T$, and the punctures cannot jump to any higher energy states. In other words, increase in temperature will not cause any re-arrangement of the punctures in the energy levels of the system. Since temperature is a parameter, which determines the average distribution of the constituents of the system in the different energy levels, we have no way to tell, whether the temperature has been increased from $T_C$ or not, and we may say as well that $T_C$ is the highest possible temperature of spacetime. If the characteristic temperature $T_C$ is both the lowest and the highest possible temperature of the spacetime, it must be {\it the only possible temperatue of spacetime} from the point of view of our observer. It is interesting to observe that if we choose the Immirzi parameter $\gamma$ such that
\begin{equation}
\gamma = \frac{\ln(2)}{\sqrt{3}\pi}\approx 0.127,
\end{equation}
then
\begin{equation}
T_C = \frac{a}{2\pi},
\end{equation}
which agrees with the Unruh temperature $T_U$ in Eq. (2.1) measured by an observer with constant proper acceleration $a$. Indeed, the number on the right hand side of Eq. (6.13) is the most popular choice for the Immirzi parameter $\gamma$. 

\section{Black Holes}

The expression in Eq. (6.14) for the temperature of spacetime with the choice (6.13) for the Immirzi parameter is the most important result obtained from our model so far. For the Schwarzschild black hole Eq. (6.14), together with Eq. (2.7) implies that the temperature of the hole from the point of view of an observer at rest with respect to the coordinates $r$, $\theta$ and $\phi$, just outside of the event horizon of the hole is:
\begin{equation}
T_C = \left(1 - \frac{2M}{r}\right)^{-1/2}\frac{M}{2\pi r^2}.
\end{equation}
Very close to the horizon we may therefore write, as an excellent approximation:
\begin{equation}
T_C = B\frac{1}{8\pi M},
\end{equation}
where
\begin{equation}
B:= \left( 1- \frac {2M}{r}\right)^{-1/2}
\end{equation}
is the blue-shift factor. According to the Tolman relation the temperature measured by a distant observer in asymptotically flat spacetime is: \cite{kaakaa}
\begin{equation}
T_\infty = \vert g_{tt}\vert^{-1/2} T,
\end{equation}
and using Eq. (2.4) we get:
\begin{equation}
T_\infty = \frac{1}{8\pi M},
\end{equation}
which agrees with the Hawking temperature
\begin{equation}
T_H := \frac{1}{8\pi M}
\end{equation}
of the Schwarzschild black hole. This is the temperature measured by a distant observer at rest with respect to the hole for the thermal radiation emitted by the hole, when the backscattering effects of the radiation from the spacetime geometry are neglected. In Appendix C we have shown that the temperature of the general, rotating, electrically charged black hole with mass $M$, electric charge $Q$ and angular momentum $J$ from the point of view of the distant observer at rest is
\begin{equation}
T_\infty = \frac{\kappa}{2\pi}
\end{equation}
where
\begin{equation}
\kappa := \frac{\sqrt{M^2 - (J/M)^2 - Q^2}}{2M[M+\sqrt{M^2 - (J/M)^2 - Q^2}] - Q^2}
\end{equation}
is the surface gravity of the hole. 

It is most gratifying to observe that our model is capable to produce, in Eq. (7.7), the standard result, usually derived by means of the quantum theory of fields in curved spacetime, for the black holes. In contrast to the results obtained in Refs. \cite{kootoo,neetoo,viitoo}, where $j_p$ was allowed to take any integer or half-integer value, and the resulting Hawking temperature was just the {\it lowest possible temperature} of the black hole, our model implies that for the black hole with parameters $M$, $J$ and $Q$ the Hawking temperature is the {\it only possible temperature} of the hole. This result of our model is consistent with the standard results obtained from the quantum field theory in curved spacetime, and it may be used as an argument for our choice to restrict the possible values of $j_p$ to $0$ and $\frac{1}{2}$.   Even more important, however, is that our model provides a microscopic explanation to the black hole radiation: When the black hole radiates, the punctures of the spin network descend from the excited states, where $j_p = \frac{1}{2}$ back to the ground state, where $j_p = 0$, and radiation is emitted. 

The really interesting results may be obtained, however, if we consider the entropic properties of black holes. In the natural units the entropy of any system is
\begin{equation}
S(\beta) = \beta E(\beta) + \ln Z(\beta),
\end{equation}
where $E(\beta)$ is the energy of the system. Employing Eqs. (5.5), (5.8) and (6.2) we may write the entropy $S$ of the black hole in terms of the parameter $z$:
\begin{equation}
S(z) = \ln(2z)\left(\frac{z}{z-1} + \frac{N}{1-z^N}\right) + \ln\left[\frac{1}{z-1}\left( 1 - \frac{1}{z^N}\right)\right].
\end{equation}
It is easy to see that 
\begin{equation}
\lim_{z\rightarrow\infty}S(z) = \ln(2),
\end{equation}
which reflects the fact that in the very low temperature just one of the punctures is in the excited state, where $j_p=\frac{1}{2}$, and $m_p$ is either $+\frac{1}{2}$ or $-\frac{1}{2}$. In other words, the ground state of the hole is two-fold degenerate. Since $N$ is assumed to be very large, we may write, in effect,
\begin{equation}
S(z) = \ln(2z)\frac{z}{z-1} - \ln(z-1),
\end{equation}
whenever $z>1$, which means that $T<T_C$. To see how the entropy behaves, when $T>T_C$, which means that $z<1$, let us write Eq. (7.10) in the form:
\begin{equation}
S(z) = N\ln(2) + \frac{z}{z-1}\ln(2z) - \ln(1-z) + \ln(1 - z^N) + \frac{z^N}{1 - z^N}N\ln(2z).
\end{equation}
When $z<1$, the last two terms on the right hand side will vanish in the large $N$ limit. Hence we have, in effect:
\begin{equation}
S(z) = N\ln(2) + \frac{z}{z-1}\ln(2z) - \ln(1-z),
\end{equation}
whenever $T>T_C$. For large $N$ the second and the third terms on the right hand side are negligible, when compared to the first term. In the special case, where $z=1$, or $T=T_C$, Eqs. (5.9), (6.3) and (7.9) imply:
\begin{equation}
S(z) = \frac{1}{2}(N+1)\ln(2) + \ln(N).
\end{equation}

Comparing Eqs. (7.12), (7.14) and (7.15) we observe that at the characteristic temperature $T_C$ the entropy of the hole performs an enormous jump. When $T<T_C$, the entropy of the hole is essentially zero, whereas when $T>T_C$, the entropy is, in effect, $N\ln(2)$. It is interesting to observe that in the large $N$ limit the entropy of the black hole is, to an excellent approximation, a {\it constant function} of the temperature $T$. In the high-temperature limit, where $z\rightarrow\frac{1}{2}$, Eq. (7.14) yields for the black hole entropy an expression $(N+1)\ln(2)$.

Eqs. (4.8) and (6.13) imply that if there are $n$ punctures in the excited state, where $j_p= \frac{1}{2}$, then the area of the stretched horizon of the black hole is of the form:
\begin{equation}
A_n = 4n\ln(2)\ell_{Pl}^2.
\end{equation}
For all practical purposes the area of the stretched horizon may be identified with the event horizon area of the hole. After the phase transition at the characteristic temperature $T_C$ all $N$ punctures at the horizon have jumped to the excited state, and the event horizon area becomes to:
\begin{equation}
A = 4N\ln(2)\ell_{Pl}^2.
\end{equation}
Comparing Eqs. (7.14) and (7.17) we may write the black hole entropy in terms of its event horizon area, in the natural units, as:
\begin{equation}
S  = \frac{1}{4}A
\end{equation}
or, in the SI units:
\begin{equation}
S = \frac{1}{4}\frac{k_Bc^3}{\hbar G}A,
\end{equation}
which is the famous {\it Bekenstein-Hawking entropy law}. It should be noted that in our model the Bekenstein-Hawking entropy law holds not only after the phase transition at the temperature $T_C$, but during the the phase transition as well. During the phase transition the temperature $T=T_C$ of the black hole is a constant, and using Eqs. (3.7) and (6.14) we find:
\begin{equation}
E = \frac{1}{4}T_C A.
\end{equation}
Employing the general result
\begin{equation}
\frac{\partial S}{\partial E} = \frac{1}{T}
\end{equation}
between the entropy $S$. energy $E$ and temperature $T$ of any system we get, using Eq. (7.20):
\begin{equation}
\frac{1}{T_C} = \frac{\partial S}{\partial E} = \frac{\partial S}{\partial A}\frac{dA}{dE} = \frac{\partial S}{\partial A}\frac{4}{T_C}.
\end{equation}
So we must have, in the natural units:
\begin{equation}
\frac{\partial S}{\partial A} = \frac{1}{4},
\end{equation}
which readily implies Eq. (7.18).

As it is well known, for a system with equal weights in its microscopic states the number of the microscopic states associated with the same macroscopic state is, in the natural units:
\begin{equation}
\Omega = e^S.
\end{equation}
Eq. (7.14) gives in the large $N$ limit, up to an insignificant multiplicative constant of order one:
\begin{equation}
\Omega = 2^N.
\end{equation}
This result means that we have managed to reduce the microscopic origin of black hole entropy in bits. In our model every puncture of the spin network on the horizon carries exactly two bits of information. Indeed, if there are $N$ punctures, and for each puncture $p$ the quantum number $m_p$ is either $+\frac{1}{2}$ or $-\frac{1}{2}$, the total number of the possible combinations of the states is $2^N$. The name {\it Bekenstein number} was coined by Wheeler in his essay for the number $N$ in Eq. (7.25). According to Wheeler it "tells the number of binary digits, the number of bits, that would be required to specify in all detail the configuration of the constituents out of which the black hole was put together." \cite{yy} In our model  the number $N$ in Eq (7.25) tells the total number of the punctures of the spin network on the horizon, and it is assumed to be a constant for the given black hole. The precise value of $N$ depends on the specific details of the gravitational collapse, which created the hole, and it varies from a hole to another. 

\section{Cosmological Constant}

The most remarkable discovery in modern cosmology is that the Universe is not only expanding, but its expansion is {\it accelerating}, and it brought to Saul Perlmutter, Brian P.  Schmidt and Adam Riess the Nobel Prize in physics of the year 2011. \cite{kaakoo} The energy needed to drive the accelerating expansion of the Universe is called {\it dark energy}. It has been estimated that around $ 70 \%$ of the total energy of the present Universe consists of the dark energy. 

The origin of the dark energy is one of the deepest mysteries of modern physics. The accelerating expansion of the Universe, and hence the dark energy, is most commonly attributed to the {\it cosmological constant}. Indeed, for a sufficiently low matter density of the Universe Einstein's field equation written with a positive cosmological constant $\Lambda$:
\begin{equation}
R_{\mu\nu} - \frac{1}{2}g_{\mu\nu}R + \Lambda g_{mu\nu} = \frac{8\pi G}{c^4}T_{\mu\nu}
\end{equation}
implies an accelerating expansion for the Universe. Unfortunately, attempts to attribute the dark energy and the accelerating expansion of the Universe to the cosmological constant bring along another problem: In the vacuum, where the energy density of the matter vanishes identically, we may write Eq. (8.1) as:
\begin{equation}
R_{\mu\nu} - \frac{1}{2}g_{\mu\nu}R = -\Lambda g_{\mu\nu}.
\end{equation}
At every point of spacetime one may introduce a locally flat Minkowski system of coordinates. In this system of coordinates $T_t^{\,\,t}$ gives the energy density of the matter, and hence we observe, with the signature $(-,+,+,+)$ for the metric, that the quantity
\begin{equation}
\rho_{vac} := \frac{c^2}{8\pi G}\Lambda
\end{equation}
may be viewed as the energy density of the vacuum. According to the standard quantum field theory the energy density of the vacuum is infinite. However, if we take into account the gravitational effects, we are suggested to think that the energy density of the vacuum is not necessarily infinite, but it has a certain  upper limit: Considering the Planck length $\ell_{Pl}$, as the smallest possible diameter of any region of space we observe (so the standard argument goes) that the maximum amount of energy inside of the region with diameter $\ell_{Pl}$ is about the same as the Planck energy 
\begin{equation}
E_{Pl} := \sqrt{\frac{\hbar c^5}{G}},
\end{equation}
since otherwise the region of space under consideration would collapse into a black hole with Schwarzschild radius $\ell_{Pl}$. Hence one obtains for the energy density of the vacuum an estimate
\begin{equation}
\rho_{vac} \sim \frac{E_{Pl}}{\ell_{Pl}^3}\sim \frac{c^7}{\hbar G^2},
\end{equation}
which gives for the cosmological constant an estimate
\begin{equation}
\Lambda \sim 8\pi\frac{c^5}{\hbar G} \sim 10^{87} s^{-2},
\end{equation}
which is around $10^{122}$ times larger than the most recent estimate
\begin{equation}
\Lambda = (1.00\pm 0.03)\times 10^{-35}s^{-2},
\end{equation}
based on the Planck data of the year 2015, for $\Lambda$. \cite{kaanee} Hence we are faced with the question that why is there a cosmological constant at all, and why is it so incredibly small.

To address the problem of the cosmological constant, consider the de Sitter spacetime, which is an empty spherically symmetric spacetime with positive cosmological constant. At least in its late stages of evolution the universe with a positive cosmological constant is, to a very good approximation, described by the de Sitter spacetime. As we noted in Section 3, the de Sitter spacetime possesses the cosmological horizon with radius $r_C = \sqrt{\frac{3}{\Lambda}}$. We also noted that in the de Sitter spacetime the stretched horizon of the black hole just outside of its event horizon must be replaced by a shrinked horizon just inside of its cosmological horizon.

Under the assumption that the proper acceleration $a$ of the observer on the shrinked horizon is a constant, no matter what may happen to the cosmological constant $\Lambda$, the results obtained in our previous discussion for the stretched horizon of the black hole will hold as such for the shrinked horizon of the de Sitter spacetime as well: The spin network has punctures on the shrinked horizon, and at each puncture the quantum number $j_p$ is either $0$ or $\frac{1}{2}$. Again, the spacetime has the characteristic temperature $T_C = \frac{a}{2\pi}$ from the point of view of the observer. Eq. (3.4) implies that for the congruence of the world lines of observers with constant coordinates $r$, $\theta$ and $\phi$ the only non-zero component of the future pointing unit tangent vector field of the congruence is
\begin{equation}
u^t = \left( 1 - \frac{\Lambda}{3}r^2\right)^{-1/2}.
\end{equation}
According to Eq. (2.2) the only non-zero component of the proper acceleration vector field $a^\mu$ is:
\begin{equation}
a^r = -\frac{\Lambda}{3} r,
\end{equation}
which means that
\begin{equation}
a = \sqrt{a_\mu a^\mu} = \left(1 - \frac{\Lambda}{3}r^2\right)^{-1/2}\frac{\Lambda}{3}r.
\end{equation}
Very close to the cosmological horizon, where $r= r_C$ we may therefore write the characteristic temperature, to a very good approximation, as:
\begin{equation}
T_C = \frac{1}{2\pi}B\sqrt{\frac{\Lambda}{3}},
\end{equation}
where 
\begin{equation}
B:= \left(1 - \frac{\Lambda}{3}r^2\right)^{-1/2}
\end{equation}
is the blue-shift factor. Employing again the Tolman relation in Eq. (7.4) we find that far from the cosmological horizon the temperature measured by the oberver is
\begin{equation}
T_0 = \frac{1}{2\pi}\sqrt{\frac{\Lambda}{3}}.
\end{equation}
The result is identical to the one obtained by Gibbons and Hawking in Ref. \cite{kaavii} by means of the methods based on the quantum theory of fields in curved spacetime. The temperature $T_0$ is the only possible temperature of the de Sitter spacetime. The result contrasts with the result obtained in Ref. [19], where $T_0$ was just the lowest possible temperature. Our result may be used as an argument for $0$ and $\frac{1}{2}$ as the only possible values of the quantum numbers $j_p$.

In our model the cosmological constant $\Lambda$ is determined by the quantum states of the spin network on its shrinked horizon. For all practical purposes we may identify the shrinked horizon area of the de Sitter spacetime with the area of its cosmological horizon. The area of the cosmological horizon of the de Sitter spacetime is:
\begin{equation}
A = 4\pi r_C^2 = 4\pi\left(\sqrt{\frac{3}{\Lambda}}\right)^2 = \frac{12\pi}{\Lambda}.
\end{equation}
Eq. (4.4) implies that with the choice (6.13) for the Immirzi parameter $\gamma$ the shrinked horizon area is
\begin{equation}
A_n = 4n\ln(2)\ell_{Pl}^2,
\end{equation}
where $n$ is the number of the punctures in the excited states, where $j_p=\frac{1}{2}$. Identifying the right hand sides of Eqs. (8.14) and (8.15) we obtain for the cosmological constant $\Lambda$, in the natural units, an expression:
\begin{equation}
\Lambda = \frac{3\pi}{n\ln(2)}
\end{equation}
or, in the SI units:
\begin{equation}
\Lambda = \frac{3\pi}{n\ln(2)}\frac{c^5}{\hbar G}.
\end{equation}

Now, if we accept the view that the cosmological constant $\Lambda$ is determined by the quantum states of the punctures of the spin network on the shrinked horizon, we must also accept the possiblity that the cosmological constant may change, when the punctures perform jumps between different quantum states. As we saw in Section 6, at the characteristic temperature $T_C$ spacetime performs a phase transition, where all $N$ punctures of the spin network on the stretched or shrinked horizon jump from the vacuum, where $j_p = 0$ to the excited state, where $j_p = \frac{1}{2}$. Under the assumption that the phase transition has been completed in the present Universe, the cosmological constant takes the form:
\begin{equation}
\Lambda = \frac{3\pi}{N\ln(2)}\frac{c^5}{\hbar G}.
\end{equation}
The number $N$, which gives the total number of the punctures of the spin network on the shrinked horizon, is the Bekenstein number of the Universe. If we choose 
\begin{equation}
N = 4.7\times 10^{122},
\end{equation}
we find that
\begin{equation}
\Lambda = 1.0\times 10^{-35}s^{-2},
\end{equation}
which agrees with the present estimate for the cosmological constant. In this sense our model provides an explanation both to the presence and the smallness of the cosmological constant. The cosmological constant is non-zero, because the Bekenstein number $N$ is finite, and it is $10^{122}$ times smaller than expected for the simple reason that the Bekenstein number is around $10^{122}$. The Universe has its own Bekenstein number in the same sense as every black hole has its own Bekenstein number, which depends on the details of the gravitational collapse during the creation of the hole. The Bekenstein number $N$ of the Universe should be viewed as a constant of nature, which determines, among other things, the magnitude of the cosmological constant. \cite{kaakuu}

\section{The Cosmic Phase Transition}

The key result of our model, inspired by Wheeler's it-from-bit proposal, is that at the characterisitic temperature $T_C = \frac{a}{2\pi}$ spacetime performs a phase transition: When spacetime is heated up, the punctures of the spin network on the stretched or shrinked horizon jump from the vacuum state, where $j_p=0$ to the excited state, where $j_p=\frac{1}{2}$, when the temperature exceeds $T_C$, and back from the excited state to the vacuum, when the spacetime is cooled down below the temperature $T_C$. The microscopic origin of the black hole radiance, for instance, is the phase transition, where the punctures descend from the excited states to the vacuum, and radiation is emitted.

It is very interesting to consider a possibility of a similar phase transition in a cosmic scale, in the whole Universe. There are good grounds to believe that the Universe began its existence in a state of very low entropy and therefore, according to the third law of thermodynamics, in a state of very low temperature. Almost immediately after its creation, however, the Universe became very hot, and its temperature exceeded the characteristic temperature $T_C$. As a consequence, the punctures of the spin network on the shrinked horizon just inside of its cosmological horizon jumped from the vacuum to the excited states, and the cosmological horizon experienced a tremendous expansion 

As we saw in Eq. (8.14), in the de Sitter spacetime the area of the cosmological constant is inversely proportional to cosmological constant $\Lambda$. Hence increasing cosmological horizon during the phase transition implies decreasing cosmological constant. In other words, it appears that the cosmological "constant" $\Lambda$ may not be constant, but it varies in time.

There is an extensive literature on the cosmological effects of the varying cosmological constant. \cite{kaaseite} In all approaches the starting point is to write Einstein's field equation (8.1), in the natural units, as:
\begin{equation}
R_{\mu\nu} - \frac{1}{2}g_{\mu\nu}R = 8\pi\tilde{T}_{\mu\nu},
\end{equation}
where
\begin{equation}
\tilde{T}_{\mu\nu} := T_{\mu\nu} - \frac{1}{8\pi}\Lambda g_{\mu\nu}.
\end{equation}
The tensor $\tilde{T}_{\mu\nu}$ involves the contribution of both the matter and the cosmological constant to the energy, momentum and stress in spacetime. The energy and the momentum described by $\tilde{T}^{\mu\nu}$ must be conserved, and therefore we must have:
\begin{equation}
\tilde{T}^{\mu\nu}_{\,\,\,;\nu} = 0
\end{equation}
for all $\mu = 0,1,2,3$, no matter, whether $\Lambda$ is conserved in time or not. Eq. (9.3) implies that whenever the cosmological constant $\Lambda$ changes in time, the energy and the momentum of the matter will also change. In other words, the energy and the momentum of the vacuum, which are proportional to the cosmological constant $\Lambda$, may be converted to the energy and the momentum of the matter, and vice versa. 

The current observational data supports the idea that the Universe is flat. In the spatially flat universe we write the line element of spacetime in terms of its scale factor $R(t)$ as:
\begin{equation}
ds^2 = -dt^2 + R^2(t)(dx^2 + dy^2 + dz^2).
\end{equation}
In Appendix D we have shown that under the assumption that the Universe is filled with perfect fluid with energy density $\rho$ and pressure $p$ Einstein's field equation implies for the scale factor $R$ an equation:
\begin{equation}
\left(\frac{\dot{R}}{R}\right)^2 = \frac{8\pi}{3}\left(\rho + \frac{1}{8\pi}\Lambda\right)
\end{equation}
which, when Eq. (9.3) is taken into account, implies an equation (see also Ref. \cite{kaaseite}):
\begin{equation}
\frac{\ddot{R}}{R} = -\frac{1}{2}(1 + 3w)\left(\frac{\dot{R}}{R}\right)^2 + \frac{w+1}{2}\Lambda.
\end{equation}
In Eqs. (9.5) and (9.6) the dot means the derivative with respect to the time $t$, whereas the number $w$ is the equation of state of the matter. In other words, between the energy density $\rho$ and the pressure $p$ there is the relationship:
\begin{equation}
w = \frac{p}{\rho}.
\end{equation}
If the Universe is assumed to be filled with frictionless, pressureless dust, we have $w=0$, whereas if the Universe is filled with homogeneous electromagnetic radiation, then $w=\frac{1}{3}$.

Eq. (9.6) is the key equation, when considering the effects of the time-dependent cosmological constant on the scale factor $R(t)$. Denoting:
\begin{equation}
y := R^{\frac{3}{2}(1+w)},
\end{equation}
Eq. (9.6) becomes simplified to the form:
\begin{equation}
\frac{\ddot{y}}{y} = \frac{3}{4}(w+1)^2\Lambda.
\end{equation}
In the special case, where $\Lambda=0$, the solution satisfying the initial condition $y(0)=0$ to this equation is
\begin{equation}
y(t) = C^{\frac{3}{2}(1+w)}t,
\end{equation}
where $C$ is a positive constant. Hence Eq. (9.8) implies:
\begin{equation}
R(t) = Ct^{\frac{2}{3(1+w)}}.
\end{equation}
If $w=0$, we get:
\begin{equation}
R(t) = Ct^{2/3},
\end{equation}
which gives the time evolution of the scale factor in the flat Friedman model of the Universe, whereas if $w=\frac{1}{3}$, we get:
\begin{equation}
R(t) = Ct^{1/2}.
\end{equation}

When considering the possible phase transition in the Universe, the crucial question is, how fast the phase transition will take place, {\it i. e.} how fast will the cosmological constant $\Lambda$ decrease in time from its Planck-size value $10^{87}s^{-2}$ to its present value $1.0\times 10^{-35}s^{-2}$. If we ignore the effects of the matter fields, then for constant $\Lambda$ both of the Eqs. (9.5) and (9.6) are satisfied by the de Sitter solution:
\begin{equation}
R(t) = R_0\exp\left(\sqrt{\frac{\Lambda}{3}}t\right),
\end{equation}
where $R_0$ is a constant. Another way to write Eq. (9.14) is:
\begin{equation}
t = \sqrt{\frac{3}{\Lambda}}\ln\left(\frac{R}{R_0}\right).
\end{equation}
If we put for the cosmological constant $\Lambda$ its presumed initial value $10^{87}s^{-2}$, and for $\frac{R}{R_0}$ the number $10^{61}$, which is the ratio of the present radius of the observable Universe, or  $10^{26}m$, to the Planck length $\ell_{Pl}\sim 10^{-35}m$, we get:
\begin{equation}
t \sim \sqrt{\frac{3}{10^{87}}}\ln(10^{61})s\sim 10^{-41}s.
\end{equation}
This means that two points originally at the Planck length distance from each other would have been torn apart to a distance greater than the present radius of the observable Universe in just $10^{-41}$ seconds. In other words, the Universe as we know it could have been created in a tiny fraction of a second. No real cosmologist could ever be persuaded to believe this, and hence we must conclude that the cosmological constant must have decreased very rapidly during the phase transition. 

At the moment we have no observational data for the rate of the decrease of the cosmological constant in the very early Universe. However, the unit of $\Lambda$ is $s^{-2}$, and hence the dimesional arguments suggest that the cosmological constant might be inversely proportional to the square of the age $t$ of the Universe. \cite{kaakasi} Indeed, if we put:
\begin{equation}
\Lambda(t) = \frac{3n^2(1+w) -2n}{1+w}\frac{1}{t^2},
\end{equation}
then Eq. (9.9) has the solution:
\begin{equation}
y(t) = C^{\frac{3}{2}(1+w)}t^{\frac{3}{2}n(1+w)},
\end{equation}
where, again, $C$ is a positive constant. Employing Eq. (9.8) we find the corresponding expression for the scale factor $R(t)$:
\begin{equation}
R(t) = Ct^n.
\end{equation}
Cosmological models, where the scale factor is proportional to some positive power of the age $t$ of the Universe are known as {\it power law cosmologies.} \cite{kaaysi} Hence we see that the power law cosmologies are consistent with the idea that the cosmological constant $\Lambda$ was inversely proportional to the square of the age $t$ of the Universe during the phase transition.

In the flat Universe the precise numerical value of the positive constant $C$ is irrelevant. The more interested, however, are we in the power $n$ of the age $t$ of the Universe. It would be very tempting to substitute for $t$ in Eq. (9.17) the present estimate for the age of the Universe, which is around 14 billion years, and for $\Lambda$ the estimate given in Eq. (8.7), and then solve $n$ from the resulting equation. Unfortunately, such an approach would be erroneous, because the age $t$ of the Universe is model-dependent. Instead of using the estimated age of the Universe as a data, when approximating $n$, we should base our estimate on the observed values of the Hubble constant $H$ and the cosmological constant $\Lambda$. The numerical values of these parameters are model-independent, and their determination may be based on direct observations. 

The Hubble constant is, by definition,
\begin{equation}
H := \frac{\dot{R}}{R},
\end{equation}
and Eq. (9.19) implies that the present age $t$ of theUniverse may be expressed in terms of the Hubble constant as:
\begin{equation}
t = \frac{n}{H}.
\end{equation}
Substituting Eq. (9.21) in Eq. (9.17) we may express the power $n$ in terms of $H$ and $\Lambda$:
\begin{equation}
n = \frac{2H^2}{(1+w)(3H^2 - \Lambda)}.
\end{equation}
In the matter-dominated universe $w=0$. Putting for $H$ its value according to the Planck data release of the year 2015, which is around $67.8\,kms^{-1}/MPc$, or $2.2\times 10^{-18}s^{-1}$ and for $\Lambda$ the estimate $1.0\times 10^{-35}s^{-2}$, we get:
\begin{equation}
n\approx 2.1.
\end{equation}
An error analysis based on the estimated errors in the measurements of $H$ and $\Lambda$ has been performed in Appendix E. Our analysis gives the power $n$ with the error bars:
\begin{equation}
n = 2.1\pm 0.2.
\end{equation}
It is very interesting that the power $n$ is so close to two. When $n\approx 2$, the expansion of the Universe is indeed accelerating. However, its present age differs from the current estimate. Putting $n=2$ and $H=2.2\times 10^{-18}s^{-1}$ in Eq. (9.21) we find that the age of the Universe should be around 30 billion years, which is more than twice the present estimate. Nevetheless, it appears that the tentative model, where the cosmological constant is assumed to be inversely proportional to the square of the age of the Universe is capable to produce the present values of the directly observed cosmological parameters.  

\section{The Binary Code of the Universe}

The idea that the cosmological constant $\Lambda$ decreases because of the phase transition taking place just inside of the cosmological horizon of the Universe has an important consequence: Matter is created out of the vacuum. For instance, if we consider a matter-dominated model of the Universe, where $w=0$ and $\Lambda$ is proportional to $t^{-2}$, Eqs. (9.5), (9.17) and (9.19) imply that the mass density $\rho$ of the Universe depends on its age $t$ as:
\begin{equation}
\rho = \frac{n}{4\pi}\frac{1}{t^2} = \frac{nc^2}{4\pi G}\frac{1}{t^2},
\end{equation}
where we have used the SI units in the last equality. If we pick up a sphere with radius $\tilde{R}$ from the flat universe, the mass of the matter inside of that sphere is, according to Eqs. (9.19) and (10.1):
\begin{equation}
M = \frac{4}{3}\pi\tilde{R}^3\rho \propto t^{3n-2}.
\end{equation}
Denoting by $M_0$ the mass inside of the sphere, when $t=t_0$, we therefore get:
\begin{equation}
M = M_0\left(\frac{t}{t_0}\right)^{3n-2}.
\end{equation}
If matter were not created nor annihilated during the expansion of the Universe, then $M$ would be a constant: Even though the mass density $\rho$ decreases in time, when the Universe expands, the total mass $M$ remains the same. Indeed, if we put $n = 2/3$, which is the case in the flat Friedman model, then $M=constant$, and Eq. (9.17) implies that $\Lambda=0$. However, if we put $n=2$, as suggested by the observations, then 
\begin{equation}
M = M_0\left(\frac{t}{t_0}\right)^4,
\end{equation}
which indicates a pretty rapid increase in the mass of the Universe. 

Our observation about the rapid increase in the amount of matter in the early Universe brings us back to the main theme of this paper: Where does the information needed for the creation of the matter come from? To begin with, let us write the FRW metric in the spatially flat Universe by means of the spherical coordinates:
\begin{equation}
ds^2  = -dt^2 + R^2(t)[dr^2 + r^2\,d\theta^2 + r^2\sin^2(\theta)\,d\phi^2].
\end{equation}
Defining the new radial coordinate 
\begin{equation}
\tilde{r} := Rr
\end{equation}
we find that
\begin{equation}
dr = d\left(\frac{\tilde{r}}{R}\right) = \frac{1}{R}\,d\tilde{r} - \frac{\tilde{r}}{R^2}\dot{R}\,dt,
\end{equation}
and Eq. (10.4) takes the form:
\begin{equation}
ds^2 = -(1-H^2\tilde{r}^2)\,dt^2 - 2H\tilde{r}\,d\tilde{r}\,dt + d\tilde{r}^2 + \tilde{r}^2\,d\theta^2 + \tilde{r}^2\sin^2(\theta)\,d\phi^2,
\end{equation}
where, again, $H := \frac{\dot{R}}{R}$ is the Hubble constant. As one may observe, spacetime has the cosmological horizon, where
\begin{equation}
\tilde{r} = \tilde{r}_C := \frac{1}{H}.
\end{equation}
However, because the Hubble constant $H$ decreases in time, the radius $r_c$ of the cosmological horizon will increase all the time. For instance, in the power law cosmology, where the scale factor $R(t)$ depends on the age $t$ of the Universe as in Eq. (9.19), $\tilde{r}_C$ may be written in the SI units as:
\begin{equation}
\tilde{r}_C = \frac{c}{n}t.
\end{equation}
Hence it follows that when $n=2$, the radius of the cosmological horizon increases with a constant speed, which is one-half of the speed of light.

When the radius of the cosmological horizon increases during the phase transition, so does the area
 \begin{equation}
A_C = 4\pi\tilde{r}_C^2
\end{equation}
of the cosmological horizon as well. For instance Eq. (10.10) implies:
\begin{equation}
A_C(t) = 4\pi\left(\frac{c}{n}\right)^2t^2.
\end{equation}
During the increase of the cosmological horizon the punctures of the spin network on the shrinked horizon jump to the excited state. Once after the phase transition has been completed, and the Universe has settled into an equilibrium, all punctures on the shrinked horizon are in the excited states. After the Universe has reached the equilibrium, the area of the cosmological horizon cannot increase any further, which implies, through Eq. (10.9), that the Hubble constant is independent of time. This means that the Universe has become, in effect, to the de Sitter universe, where the scale factor increases exponentially in the time $t$. As it was mentioned in Section 3, the quantity
\begin{equation}
E = \frac{a}{8\pi}A
\end{equation}
may be identified as the energy of the de Sitter spacetime from the point of view of an observer with constant proper acceleration $a$, just inside of the cosmological horizon. The area $A$ of the shrinked horizon may be equated, for all practical purposes, with the area of the cosmological horizon. A chain of reasoning identical to the one carried out for black holes in Section 7 produces the result that de Sitter spacetime possesses entropy which, as an excellent approximation, may be written as:
\begin{equation}
S = \frac{1}{4}A,
\end{equation}
which is identical to the Bekenstein-Hawking entropy law for black holes.

The fact that we may associate the Bekenstein-Hawking entropy with the de Sitter universe, which presumably is the final equilibrium state of the Universe, reveals an interesting connection with Wheeler's it-from-bit proposal and the structure of the Universe: In the same way as the Bekenstein-Hawking entropy of a black hole tells the amount of information needed for the creation of the hole, so is one tempted to interpret the Bekenstein-Hawking entropy of the de Sitter universe as the amount of information needed for the creation of the matter out of the vacuum during the phase transition in the observable part of the Universe. In other words, all information about the properties of the observable part of the Universe is ultimately encoded to the properties of its cosmological horizon. If our interpretation turns out to be correct, the model used in this paper allows us to reduce this information in bits: At every puncture $p$ of the spin network on the shrinked horizon the quantum number $m_p$ may take exactly two values, which are $\frac{1}{2}$ and $-\frac{1}{2}$, and therefore every puncture carries exactly one bit of information. For every possible state of the observable Universe there exists a unique combination of the quantum numbers $m_p$ at the punctures. In this sense the combination of the quantum numbers $m_p$ gives the {\it binary code} of the Universe as such as it appears to us. Presumably, the number of the punctures on the shrinked horizon is around $10^{122}$ or so, and hence the number of the possible binary codes is around $2^{10^{122}}$. Out of these $2^{10^{122}}$ possible binary codes exactly one represents the correct final state of the Universe. So we see that it is indeed possible to realize Wheeler's grand vision of the reduction of everything - literally everything - in bits.

\section{Concluding Remarks}

In this paper we have considered a possiblity to realize at least some aspects of Wheeler's it-from-bit proposal by means of loop quantum gravity. To this end we considered a model, where the only possible values taken by the quantum numbers $j_p$ at the punctures $p$ of the spin network on the spacelike two-surfaces of spacetime were assumed to be $0$ and $\frac{1}{2}$. When $j_p=0$, the puncture is in the vacuum, and it does not contribute to the area of the two-surface, whereas when $j_p=\frac{1}{2}$, the puncture is in an excited state, and the possible values taken by the associated quantum number $m_p$ are $+\frac{1}{2}$ and $-\frac{1}{2}$. In this sense the punctures in our model are analogous to the spin $\frac{1}{2}$ particles, which according to quantum field theory may be created out of the vacuum, and also annihilated, and which have spin, which is either up or down. Since at each puncture $p$ in an excited state the quantum number $m_p$ is either $+\frac{1}{2}$ or $-\frac{1}{2}$, each puncture carries exactly one bit of information. We applied our model for spacetimes with horizons. The horizon of spacetime may be, for instance, the event horizon of a black hole, or the cosmological horizon of an expanding universe. With some very simple assumptions concerning the statistics of the punctures we managed to obtain an explicit expression for the partition function of spacetime from the point of view of an observer with constant proper acceleration just outside of the event horizon of a black hole or just inside of he cosmological horizon of the Universe. Among other things, our partition function implied, with an appropriate choice of the Immirzi parameter, the Bekenstein-Hawking entropy law for the event horizon of the black hole, and an analogous expression for the entropy of the cosmological horizon of the Universe.

The key implication of the partition function in our model was an existence of a {\it phase transition} at a temperature, which happens to agree, from the point of view of our observer with constant proper acceleration $a$, with the Unruh temperature $T_U = \frac{a}{2\pi}$ measured by the observer: When the temperature $T$ of the spacetime is less than $T_U$, the punctures of the spin network on the event horizon of a black hole are effectively in the vacuum, where $j_p=0$, and there is no black hole. When $T=T_U$, however, the punctures jump from the vacuum to the excited states, where $j_p=\frac{1}{2}$. In this sense the Unruh temperature $T_U$ may be regarded as the lowest possible temperature of the black hole from the point of view of our observer. However, once after all of the punctures have reached the excited state, where $j_p=\frac{1}{2}$, they cannot be excited any further even if we increased the temperature, and therefore the Unruh temperature $T_U$ may be regarded as the highest possible temperature of the hole as well. It turns out that the Unruh temperature of our observer corresponds to the Hawking temperature $T_H$ measured by a distant observer at rest with respect to the hole, and hence we may conclude that the Hawking temperature $T_H$ is the only possible temperature of the black hole from the point of view of the faraway observer: The black hole emits radiation with the characteristic temperature $T_H$, and when the black hole radiates, the punctures jump from the excited states, where $j_p=\frac{1}{2}$ back to the vacuum, where $j_p=0$. 

Of particular interest are the cosmological effects implied by the phase transition in our model. It is very likely that the Universe began its existence in a state of very low entropy and therefore, according to the third law of thermodynamics, in a state of very low temperature. Almost immediately after its creation, however, its temperature became very high and the punctures of the spin network on the shrinked horizon jumped from the vacuum to the excited states. As a consequence, the  area of the cosmological horizon went through a dramatic increase. We considered a possiblity that at the beginning the Universe was a de Sitter universe, which is an empty universe with a positive cosmological constant. In the de Sitter universe the area of the cosmological horizon is inversely proportional to the cosmological constant, and hence increasing area of the cosmological horizon means decreasing cosmological constant. In other words, the cosmological "constant" $\Lambda$ is not necessarily a constant, but it may decrease in time. Our model provides a possible solution to the problem of the cosmological constant: At the beginning the cosmological constant had a value expected on dimensional grounds, or around $10^{87}s^{-2}$. After the phase transition, however, the cosmological constant settled to its present value, which is around $10^{122}$ orders of magnitude less, or $10^{-35}s^{-2}$. Indeed, if one assumes that the number $N$ of the punctures on the shrinked horizon is around $10^{122}$, our model predicts the observed value for the cosmological constant. In other words, the reason why the present value of the cosmological constant is around $10^{122}$ times less than expected is, quite simply, that there are $10^{122}$ punctures of the spin network on the shrinked horizon of the cosmological horizon. In general, the amount of entropy carried by a horizon with $N$ punctures is, to an excellent approximation,
\begin{equation}
S = N\ln(2).
\end{equation}
This result means that a horizon with $N$ punctures carries $N$ bits of information. Hence the cosmogical horizon carries around $10^{122}$ bits of information.

At the moment we do not have enough observational data to be able to decide, whether the possible phase transition making the cosmological constant to decrease in time is still going on, and if so, at which rate. In this paper we considered tentatively the so-called power law cosmology, where the Universe is assumed to be spatially flat and the scale facor $R(t)$ of the Universe depends on its age $t$ as:
\begin{equation}
R(t) = Ct^n,
\end{equation}
where $C$ is a constant. We found that in the matter-dominated Universe the cosmological constant $\Lambda$ is inversely proportional to the square of the age of the Universe, whenever $n>2/3$. Based on the Planck data of the year 2015 we estimated that
\begin{equation}
n=2.1\pm 0.2.
\end{equation}
This choice of $n$, when combined with the Planck data, gave around 30 billion years for the estimated age of the Universe. The result is about two times the currently accepted estimate, which is around 14 billion years. However, it should be remembered that the present estimate for the age of the Universe is based on an assumption that the cosmological constant does not change in time. If the cosmological constant is assumed to change in time, the resulting  estimate for the age of the Universe will be different.

Unfortunately, decreasing cosmological constant implies non-conservation of matter and energy. More precisely, the energy of the vacuum, which is represented by the cosmological constant $\Lambda$, is converted to the energy of the matter. As a consequence, matter and energy are created all the time out of the vacuum. In our model the Universe was originally empty, with a huge Planck-size cosmological constant. However, when the cosmological constant decreased because of the phase transition at the cosmological horizon, the Universe became filled with matter. The matter could not have been created without information. In our model the information needed for the creation of the matter in the observable part of the Universe is encoded to the cosmological horizon. More precisely, the information is encoded to the punctures of the spin network at the shrinked horizon just inside of the cosmological horizon. At each puncture $p$ the quantum number $m_p$ is either $+\frac{1}{2}$ or $-\frac{1}{2}$, and in this sense the combination of the values taken by $m_p$ at the punctures gives the "binary code" of the Universe. It is this binary code, where the information about everything in the observable Universe resides. Hence our model allows us to reduce everything in the observable Universe in bits, as requested by Wheeler. 

As we have seen, our model meets with some success in the sense that it makes a connection with Wheeler's it-from-bit proposal, explains the microscopic origin of the black hole entropy, and provides a possible reason for the existence and the smallness of the cosmological constant. Nevertheless, there is no doubt that the model deserves some critique as well. For instance, one might consider the fact that the final value taken by the cosmological constant after the phase transition depends on the number  $N$ of the punctures of the spin network at the shrinked horizon of the cosmological horizon as a disadvantage of our model. In a sense it appears as if we had just replaced the arbitrary cosmological constant $\Lambda$ by a new arbitrary constant $N$. Such a conclusion, however, would be erroneous: The number $N$ of the punctures tells the number of the bits of information in the system bounded by a horizon, no matter whether that system is a black hole, or the observable part of the Universe, and it is proportional to the area of the horizon. Each system carries a specific amount of bits of information, which Wheeler called as the Bekenstein number of the system. For a black hole, for instance, the Bekenstein number depends on the properties of the matter, which collapsed into the hole. It is still unknown, what determines the Bekenstein number of the Universe, and we must accept that number as a fundamental constant of nature.

The most importat concrete prediction of our model is the spontaneous creation of matter out of the vacuum during the cosmic phase transition, where the cosmological constant decreased. It would be very interesting to consider the possible observational implications of this prediction, especially if the phase transition were still going on in the present Universe. Of particular interest would be a study of its effects on the galaxy rotation curves and other phenomena usually attributed to dark matter.

The famous essay written by Wheeler is very deep, and in this paper we have been able to capture just a  one specific topic of that essay. For instance, the quantum-mechanical interplay between the observer and the observed, which is one of the main themes of Wheeler's essay, was left completely out of the discussion in our paper. However, it is to be hoped that this paper has been able to convey at least something of the general spirit of Wheeler's essay, and to some extent of his whole work in physics which, in very short terms, may be described as a strive for the fundamental simplicity. Indeed, in this paper we have attempted to realize Wheeler's it-from-bit proposal by means of the simplest possible model, which is consistent with everything else we know about physics. Some day, perhaps, the physicists will be able to say, in Wheeler's words: "Oh, how could it have been otherwise! How could we all have been so blind so long!" \cite{yy}

\appendix

\section{The Derivation of Eq. (3.2)}

When written the in the Boyer-Lindquist coordinates the line element of spacetime involving the Kerr-Newman black hole takes, in the natural units, the form: \cite{kolkyt}
\begin{equation}
\begin{split}
ds^2 = &-\frac{\Delta - b^2\sin^2(\theta)}{\Sigma}\,dt^2 - \frac{2b\sin^2(\theta)(r^2 + b^2 - \Delta)}{\Sigma}\,dt\,d\phi\\
 &+\frac{(r^2 + b^2)^2 - \Delta b^2\sin^2(\theta)}{\Sigma}\sin^2(\theta)\,d\phi^2 + \frac{\Sigma}{\Delta}\,dr^2 + \Sigma\,d\theta^2,
\end{split}
\end{equation}
where
\begin{subequations}
\begin{eqnarray}
\Sigma :&=& r^2 + b^2\cos^2(\theta),\\
\Delta :&=& r^2 + b^2 + Q^2 - 2Mr.
\end{eqnarray}
\end{subequations}
In these equations, $M$ is the ADM mass of the hole, $Q$ its electric charge, and $b$ is the angular momentum per unit mass. The black hole has the {\it horizons}, when
\begin{equation}
\Delta = 0,
\end{equation}
which means that
\begin{equation}
r = r_{\pm} := M \pm \sqrt{M^2 - b^2 - Q^2}.
\end{equation}
The spacelike two-surface, where $r = r_+$ is the event horizon of the hole. In addition to the horizons, however, the Kerr-Newman black hole also has the so-called {\it ergosphere}, where $r \ge r_+$, and
\begin{equation}
\Delta - b^2\sin^2(\theta) \le 0,
\end{equation}
which means that
\begin{equation}
r_+ < r \le M + \sqrt{M^2 - b^2\cos^2(\theta) - Q^2}.
\end{equation}
When $\theta = 0$ or $\theta = \pi$, the corresponding points on the outer boundary of the ergosphere lie on the event horizon, whereas otherwise those points lie outside of the event horizon.

   The problem with the ergosphere is that in the ergosphere the coordinate $t$ ceases to be timelike. To investigate this problem, let us write the Kerr-Newman metric in the ADM form:
\begin{equation}
ds^2 = -(N^2 - N^\phi N_\phi)\,dt^2 + 2N_\phi\,dt\,d\phi +q_{rr}\,dr^2 + q_{\theta\theta}\,d\theta^2 + q_{\phi\phi}\,d\phi^2,
\end{equation}
where we have denoted:
\begin{subequations}
\begin{eqnarray}
q_{rr} :&=& \frac{\Sigma}{\Delta},\\
q_{\theta\theta} :&=& \Sigma,\\
q_{\phi\phi} :&=& \frac{(r^2 + b^2)^2 - \Delta b^2\sin^2(\theta)}{\Sigma}\sin^2(\theta),\\
N_\phi :&=& - \frac{b\sin^2(\theta)(r^2 + b^2 - \Delta)}{\Sigma},\\
N^\phi :&=& - \frac{b(r^2 + b^2 - \Delta)}{(r^2 + b^2)^2 - \Delta b^2\sin^2(\theta)},\\
N^2 :&=& \frac{\Delta\Sigma}{(r^2 + b^2)^2 - \Delta b^2\sin^2(\theta)}.
\end{eqnarray}
\end{subequations}
In Eq. (A7) $N$ is the lapse, $N^\phi$ is the shift, and the quantities $q_{rr}$, $q_{\theta\theta}$ and $q_{\phi\phi}$ are the components of the metric tensor induced on the three-surface, where $t = constant$. Another way to write Eq. (A7) is:
\begin{equation}
ds^2 = -N^2\,dt^2 + q_{\phi\phi}(d\phi + N^\phi\,dt)^2 + q_{rr}\,dr^2 + q_{\theta\theta}\,d\theta^2,
\end{equation}
and so we observe that if we keep the coordinates $t$, $r$ and $\theta$ unchanged, but define a new coordinate
\begin{equation}
\phi' := \phi + N^\phi t,
\end{equation}
the line element takes the form:
\begin{equation}
\begin{split}
ds^2 = &-N^2\,dt^2 + \left\lbrack q_{rr} + q_{\phi\phi} \left(\frac{\partial N^\phi}{\partial r}\right)^2t^2\right\rbrack\,dr^2\\
&+ \left\lbrack q_{\theta\theta} + q_{\phi\phi} \left(\frac{\partial N^\phi}{\partial \theta}\right)^2 t^2\right\rbrack\,d\theta^2+ q_{\phi\phi}\,d{\phi'}^2.
\end{split}
\end{equation}
In this slicing of spacetime the coordinate $t$ is timelike, even when we are in the ergosphere. Another attractive feature of our slicing, where the three-surfaces $t = constant$  are used as the spacelike hypersurfaces of spacetime is that the coordinate curves are orthogonal. Unfortunately, the metric is no more static, but it depends on the time $t$. However, it is easy to see that close to the horizon, where $\Delta$ tends to zero, the time-dependent terms of the metric become irrelevant: In the term proportional to $dr^2$ $q_{rr}$ is proportional to $\Delta^{-1}$, whereas $\frac{\partial N^\phi}{\partial r}$ is, in the leading approximation for small $\Delta$, proportional to $\Delta^0$. In the term proportional to $d\theta^2$, in turn, $q_{\theta\theta}$ is proportional to $\Delta^0$, whereas $\frac{\partial N^\phi}{\partial \theta}$ is, in the leading approximation, proportional to $\Delta$. Hence we may neglect the time-dependent terms, when we are close to the horizon. 

The only non-zero component of the future pointing unit tangent vector field $u^\mu$ of the congruence of the world lines of observers at rest with respect to the coordinates $r$, $\theta$ and $\phi'$ in the slicing introduced in Eq. (A11) is
\begin{equation}
u^t = \frac{1}{N}.
\end{equation}
The only non-vanishing components of the proper acceleration vector field
\begin{equation}
a_\mu := u^\alpha u_{\mu;\alpha}
\end{equation}
of this congruence are:
\begin{subequations}
\begin{eqnarray}
a_r &=& \frac{\partial\ln(N)}{\partial r},\\
a_\theta &=& \frac{\partial\ln(N)}{\partial \theta},
\end{eqnarray}
\end{subequations}
and Eq. (A8f) implies:
\begin{equation}
\begin{split}
a_r =& \frac{1}{2}\frac{1}{\Delta}\frac{\partial \Delta}{\partial r} + \frac{1}{2}\frac{1}{\Sigma}\frac{\partial \Sigma}{\partial r}\\
& - \frac{1}{2}\frac{1}{(r^2 + b^2)^2 - \Delta a^2\sin^2(\theta)}\frac{\partial}{\partial r}[(r^2 + b^2)^2 -\Delta b^2\sin^2(\theta)],
\end{split}
\end{equation}
and
\begin{equation}
a_\theta = \frac{1}{2}\frac{1}{\Sigma}\frac{\partial \Sigma}{\partial\theta} + \frac{\Delta b^2\sin(\theta)\cos(\theta)}{(r^2 + b^2)^2 - \Delta b^2\sin^2(\theta)}.
\end{equation}
Close to the horizon, where $\Delta = 0$, the first terms on the right hand sides of Eqs. (A15) and (A16) will dominate. So we may write, just outside of the event horizon:
\begin{subequations}
\begin{eqnarray}
a_r &=& \frac{1}{2}\frac{1}{\Delta}\frac{\partial \Delta}{\partial r} + O(\Delta^0),\\
a_\theta &=& \frac{1}{2}\frac{1}{\Sigma}\frac{\partial \Sigma}{\partial \theta} + O(\Delta^1),
\end{eqnarray}
\end{subequations}
where $O(\Delta^0)$ and $O(\Delta^1)$, respectively, denote the terms proportional to the zeroth and the first powers of $\Delta$. The norm of the proper acceleration vector field $a_\mu$ is:
\begin{equation}
a := \sqrt{a_\mu a^\mu}
\end{equation}
and Eqs. (A8a), (A8b), (A11), (A17a) and (A17b) imply:
\begin{equation}
a = \frac{1}{2}\frac{1}{\sqrt{\Sigma\Delta}} \frac{\partial\Delta}{\partial r} + O(\Delta^0).
\end{equation}
This is the proper acceleration measured by an observer with constant coordinates $r$, $\theta$ and $\phi'$ for a particle in a free fall just outside of the event horizon.

  We now consider such spacelike two-surface just outside of the event horizon of the Kerr-Newman black hole, where $t=constant$ and, at the same time,
\begin{equation}
a = constant
\end{equation}
in every point of the two-surface. For the sake of simplicity we shall call the spacelike two-surface with these properties as the {\it stretched horizon} of the Kerr-Newman black hole. When the black hole radiates, the parameters $M$, $b$ and $Q$, as well as the coordinates $r$ and $\theta$ of any point of the stretched horizon will change, but in such a way that the proper acceleration $a$, as such as it was given in Eq. (A19), is kept as a constant. Hence we must have:
\begin{equation}
da = \frac{\partial a}{\partial M}\,dM + \frac{\partial a}{\partial b}\,db + \frac {\partial a}{\partial Q}\,dQ + \frac{\partial a}{\partial r}\,dr + \frac{\partial a}{\partial \theta}\,d\theta  = 0,
\end{equation}
where $dM$, $db$, $dQ$, $dr$ and $d\theta$, respectively, are the changes taken by $M$, $b$, $Q$, $r$ and $\theta$.
Employing Eq. (A19) we find, using the identities:
\begin{equation}
\frac{\partial \Delta}{\partial \theta} = \frac{\partial \Sigma}{\partial M} = \frac{\partial \Sigma}{\partial Q} = \frac{\partial^2 \Delta}{\partial b\partial r} = \frac{\partial^2\Delta}{\partial Q\partial r} = 0,
\end{equation}
the result:
\begin{equation}
\begin{split}
&\frac{\partial \Delta}{\partial M}\,dM + \frac{\partial \Delta}{\partial b}\,db + \frac{\partial \Delta}{\partial Q}\,dQ + \frac{\partial \Delta}{\partial r}\,dr\\
&=  \frac{\Delta}{\Sigma}\left(\frac{\partial \Delta}{\partial r}\right)^{-1}\bigg[2\Sigma\frac{\partial^2 \Delta}{\partial M\partial r}\,dM - \frac{\partial \Sigma}{\partial a}\frac{\partial \Delta}{\partial r}\,db\\
 &\,\,\,\,\,+ (2\Sigma\frac{\partial^2 \Delta}{\partial r^2} - \frac{\partial \Sigma}{\partial r}\frac{\partial \Delta}{\partial r})\,dr - \frac{\partial \Sigma}{\partial \theta}\frac{\partial \Delta}{\partial r}\,d\theta\bigg].
\end{split}
\end{equation}
Since $\Delta = 0$ at the event horizon, we observe that close to the event horizon our stretched horizon has the property that when  $M$, $b$, $Q$ and $r$ take the changes $dM$, $db$, $dQ$ and  $dr$, respectively, then:
\begin{equation}
\frac{\partial \Delta}{\partial M}\,dM + \frac{\partial \Delta}{\partial b}\,db + \frac{\partial \Delta}{\partial Q}\,dQ + \frac{\partial \Delta}{\partial r}\,dr = 0.
\end{equation}
However, on the left hand side of this equation we have nothing else, but the total differential $d\Delta$ of $\Delta$. Hence we may write Eq. (A24) as:
\begin{equation}
d\Delta = 0.
\end{equation}
This means that a stretched horizon originally close to the event horizon will stay close to the event horizon. In this sense our stretched horizon is well chosen.

    The concept of energy plays a key role in the thermodynamical investigation of all systems. A concept of energy frequently used in general relativity is the so-called {\it Brown-York energy} \cite{kaatoo}
\begin{equation}
E_{BY} := -\frac{1}{8\pi}\oint_{S^{(2)}}(k-k_0)\,d\mathcal{A}.
\end{equation}
In Eq. (A25) $k$ is the trace of the exterior curvature tensor induced on a closed, spacelike two-surface $S^{(2)}$ embedded into a spacelike hypersurface of spacetime, where the time coordinate $t = constant$. $k_0$ is the trace of the exterior curvature tensor, when the two-surface $S^{(2)}$ has been embedded in flat spacetime. $d\mathcal{A}$ is the area element on the two-surface, and we have integrated over the whole two-surface.

     The Brown-York energy may be understood as the energy of the gravitational field inside of the closed two-surface $S^{(2)}$. In this paper we shall take the two-surface $S^{(2)}$ to be the stretched horizon, where the proper acceleration $a = constant$, just outside of the event horizon of the Kerr-Newman black hole. Keeping the parameters $M$, $b$ and $Q$ as constants on that two-surface the changes $dr$, $d\theta$ and $d\phi'$ in the coordinates $r$, $\theta$ and $\phi'$, when we move from a one to another point on the stretched horizon satisfy an equation:
\begin{equation}
\frac{\partial a}{\partial r}\,dr + \frac{\partial a}{\partial\theta}\,d\theta + \frac{\partial a}{\partial \phi'}\,d\phi' = 0.
\end{equation}
According to Eqs. (A18) and (A23) we have: 
\begin{equation}
dr = \left[-\frac{\Delta}{\Sigma}\left(\frac{\partial \Delta}{\partial r}\right)^{-1}\frac{\partial \Sigma}{\partial\theta} + O(\Delta^2)\right]\,d\theta,
\end{equation}
where $O(\Delta^2)$ denotes the terms proportional to the second or higher powers of $\Delta$. Hence we observe that close to the event horizon $dr$ is significantly smaller than $d\theta$, whereas $d\phi'$ does not depend on $dr$ at all. This means that close to the event horizon our stretched horizon is, in practice, orthogonal to the coordinate curves associated with the Boyer-Lindquist radial coordinate $r$. In other words, the coordinate $r$ is, as an excellent approximation, a constant on the stretched horizon.

    On a spacelike two-surface of the Kerr-Newman spacetime, where both of the coordinates $r$ and $t$ are constants, the only non-zero components of the exterior curvature tensor induced on the two-surface are, in the leading approximation for small $\Delta$:
\begin{subequations}
\begin{eqnarray}
k_{\theta\theta} &=& - \left(\frac{\Sigma}{\Delta}\right)^{1/2}\Gamma_{\theta\theta}^r = \frac{1}{2}\left(\frac{\Delta}{\Sigma}\right)^{1/2}\frac{\partial q_{\theta\theta}}{\partial r},\\
k_{\phi\phi} &=& -\left(\frac{\Sigma}{\Delta}\right)^{1/2}\Gamma_{\phi\phi}^r = \frac{1}{2}\left(\frac{\Delta}{\Sigma}\right)^{1/2}\frac{\partial q_{\phi\phi}}{\partial r},
\end{eqnarray}
\end{subequations}
and its trace is:
\begin{equation}
k = k_\theta^{\,\,\theta} + k_\phi^{\,\,\phi} = \frac{1}{2}\left(\frac{\Delta}{\Sigma}\right)^{1/2}\frac{\partial}{\partial r}\ln(q_{\theta\theta}q_{\phi\phi}).
\end{equation}
In the flat spacetime $M = b = Q = 0$, which implies:
\begin{subequations}
\begin{eqnarray}
q_{\theta\theta} &=& r^2,\\
q_{\phi\phi} &=& r^2\sin^2(\theta),\\
\Delta &=& r^2,\\
\Sigma &=& r^2.
\end{eqnarray}
\end{subequations}
So we find that in the flat spacetime the trace of the exterior curvature tensor takes the form:
\begin{equation}
k_0 = \frac{2}{r^2}.
\end{equation}
Because the area element on the spacelike two-surface, where $r = constant$ is:
\begin{equation}
d\mathcal{A} = \sqrt{q_{\theta\theta}q_{\phi\phi}}\,d\theta\,d\phi,
\end{equation}
the Brown-York energy becomes to:
\begin{equation}
E_{BY} = -\frac{1}{8\pi}\int_0^{2\pi}d\phi \int_0^\pi d\theta\left[\left(\frac{\Delta}{\Sigma}\right)^{1/2} \frac{\partial}{\partial r}\sqrt{q_{\theta\theta}q_{\phi\phi}}\right] + \frac{1}{4\pi}\frac{1}{r}A,
\end{equation}
where
\begin{equation}
A := \int_0^{2\pi} d\phi \int_0^\pi d\theta\sqrt{q_{\theta\theta}q_{\phi\phi}}
\end{equation}
is the area of the two-surface, where $r = constant$

   Consider now what happens to the Brown-York energy, when the parameters $M$, $b$ and $Q$ take on infitesimal changes $dM$, $db$ and $dQ$, respectively, and we are close to the event horizon, where $\Delta = 0$. The resulting change in the Brown-York energy is:
\begin{equation}
\begin{split}
dE_{BY} &= -\frac{1}{8\pi}\int_0^{2\pi} d\phi \int_0^\pi d\theta\left[\frac{1}{2}\frac{1}{\sqrt{\Delta\Sigma}}\left(\frac{\partial \Delta}{\partial M}\,dM + \frac{\partial \Delta}{\partial b}\,db + \frac{\partial \Delta}{\partial Q}\,dQ\right) + O(\Delta^0)\right]\\
&\,\,\,\,\,\times\frac{\partial}{\partial r}\sqrt{q_{\theta\theta}q_{\phi\phi}},
\end{split}
\end{equation}
where $O(\Delta^0)$ denotes the terms proportional to the zeroth or higher powers of $\Delta$. Eq. (A24) implies:
\begin{equation}
\frac{\partial \Delta}{\partial M}\,dM + \frac{\partial \Delta}{\partial b}\,db + \frac{\partial \Delta}{\partial Q}\,dQ = - \frac{\partial \Delta}{\partial r}\,dr,
\end{equation}
and using Eq. (A19) we get:
\begin{equation}
dE_{BY} = \frac{1}{8\pi}\int_0^{2\pi} d\phi \int_0^\pi d\theta a\frac{\partial}{\partial r}\sqrt{q_{\theta\theta}q_{\phi\phi}}\,dr + O(\Delta^0).
\end{equation}
According to Eq. (A34) we may write the change in the area $A$ of the two-surface $r = constant$ as:
\begin{equation}
dA = \int_0^{2\pi} d\phi \int_0^\pi d\theta \frac{\partial}{\partial r}\sqrt{q_{\theta\theta}q_{\phi\phi}}\,dr.
\end{equation}
Hence it follows that we may write the change in the Brown-York energy measured by an observer on the stretched horizon $a = constant$ in terms of the change $dA$ of the stretched horizon area as:
\begin{equation}
dE_{BY} = \frac{a}{8\pi}\,dA,
\end{equation}
which is Eq. (3.2). During the creation of the Kerr-Newman black hole by means of the gravitational collapse the area of the stretched horizon $a = constant$ of the hole increases from zero to $A$. So we find that the Brown-York energy of the Kerr-Newman black hole from the point of view of an observer residing on the stretched horizon takes the form:
\begin{equation}
E_{BY} = \frac{a}{8\pi}A.
\end{equation}
For all practical purposes we may identify the stretched horizon area $A$ with the event horizon area
\begin{equation}
A_H = 4\pi(r_+^2 + b^2)
\end{equation}
of the Kerr-Newman black hole. Our expression in Eq. (A41) for the Brown-York energy of the Kerr-Newman black hole is identical to the one obtained in Ref. \cite{kootoo} for the Schwarzschild black hole, and in Ref. \cite{neetoo} for the Reissner-Nordstr\"om black hole.

\section{The Derivation of Eqs. (6.3) and (6.9)}

We found in Eqs. (5.8) and (5.9) that
\begin{equation}
Z(z) = \frac{1}{z-1}\left(1 - \frac{1}{z^N}\right),
\end{equation}
whenever $z\ne 1$, and
\begin{equation}
Z(z) = N,
\end{equation}
when $z=1$. Eqs. (5.5) and (6.1) imply that
\begin{equation}
E(z) = -\frac{Z'(z)}{Z(z)}\frac{dz}{d\beta} = -\ln(2)T_C\frac{Z'(z)}{Z(z)}z.
\end{equation}
Now, we have:
\begin{equation}
Z'(1) = \lim_{h\rightarrow 0}\frac{Z(1+h) - Z(1)}{h},
\end{equation}
and Eqs. (B1) and (B2) imply:
\begin{equation}
Z'(1) = \lim_{h\rightarrow 0}\frac{1}{h}\left\lbrace\frac{1}{h}\left[1 - \frac{1}{(1+h)^N}\right] - N\right\rbrace.
\end{equation}
Employing Newton's binomial formula:
\begin{equation}
(1+x)^N = 1 + nx + \frac{n(n-1)}{2!}x^2 + \frac{n(n-1)(n-2)}{3!}x^3+\cdots
\end{equation}
we get:
\begin{equation}
\frac{1}{(1+h)^N} = 1 - Nh + \frac{N(N+1)}{2!}h^2 - \frac{N(N+1)(N+2)}{3!}h^3 + \cdots,
\end{equation}
and therefore:
\begin{equation}
Z'(1) = -\frac{N(N+1)}{2!}.
\end{equation}
Hence Eqs. (B2) and (B3) imply:
\begin{equation}
E(1) = \frac{1}{2}(N+1)T_C\ln(2),
\end{equation}
which is Eq. (6.3).

To obtain Eq. (6.9) we note first that
\begin{equation}
\frac{dE}{dT} = \frac{dE}{dz}\frac{dz}{dT} = -E'(z)z\ln(2)\frac{T_C}{T^2}.
\end{equation}
Because 
\begin{equation}
E'(1) = \lim_{h\rightarrow 0}\frac{E(1+h) - E(1)}{h},
\end{equation}
Eqs. (6.2) and (6.3) imply:
\begin{equation}
E'(1) = \lim_{h\rightarrow 0}\frac{1}{h}\left[\frac{1+h}{h} + \frac{N}{1-(1+h)^N} - \frac{1}{2}(N+1)\right]T_C\ln(2).
\end{equation}
Again, using Eq. (B6) we find:
\begin{equation}
\frac{N}{ 1 - (1+h)^N} = -\frac{1}{h}\frac{1}{1 + \frac{N-1}{2!}h + \frac{(N-1)(N-2)}{3!}h^2 + \cdots},
\end{equation}
which may be written as:
\begin{equation}
\begin{split}
\frac{N}{1 - (1+h)^N} &= -\frac{1}{h}\bigg[1 - \frac{N-1}{2}h\\
 &+ \frac{3(N-1)^2 - 2(N-1)(N-2)}{12}h^2 + O(h^3)\bigg],
\end{split}
\end{equation}
where $O(h^3)$ denotes the terms proportional to the third or higher powers of $h$. Substituting Eq. (B14) in Eq. (B12) we get:
\begin{equation}
E'(1) = -\frac{3(N-1)^2 - 2(N-1)(N-2)}{12}T_C\ln(2).
\end{equation}
Denoting by $O(N)$ the terms proportional to the first or lower powers of the large number $N$ we obtain, by means of Eq. (B10):
\begin{equation}
\frac{dE}{dT}\bigg\vert_{T=T_C} = \frac{1}{12}N^2[\ln(2)]^2 + O(N),
\end{equation}
which is Eq. (6.9).

\section{The Derivation of Eq. (7.7)}

According to Eq. (6.14) and (A19) the characteristic temperature of the Kerr-Newman black hole from the point of view of an observer close to the event horizon, at rest with respect to the coordinates $r$, $\theta$ and $\phi'$ is.
\begin{equation}
T_C = \frac{1}{4\pi}\left[\frac{1}{\sqrt{\Sigma\Delta}}\frac{\partial\Delta}{\partial r} + O(\Delta^0)\right],
\end{equation}
where $O(\Delta^0)$ denotes the terms proportional to the zeroth and higher powers of $\Delta$. Employing Eqs. (A2b), (A8f) and (A11), together with the Tolman relation in Eq. (7.4) we find that the temperature from the point of view of the distant observer is:
\begin{equation}
T_\infty = \lim_{r\rightarrow r_+}\left(\frac{N}{2\pi}\frac{r - M}{\sqrt{\Sigma\Delta}}\right) = \frac{1}{2\pi}\lim_{r\rightarrow r_+}\left[\frac{r - M}{\sqrt{(r^2 + b^2)^2 - \Delta b^2\sin^2(\theta)}}\right],
\end{equation}
where $b:=J/M$. Using Eq. (A4) and denoting:
\begin{equation}
\kappa := \frac{\sqrt{M^2 - (J/M)^2 - Q^2}}{2M[M+ \sqrt{M^2 - (J/M)^2 - Q^2}] - Q^2}
\end{equation}
we get:
\begin{equation}
T_\infty = \frac{\kappa}{2\pi},
\end{equation}
which is Eq. (7.7).

\section{The Derivation of Eq. (9.6)}

With the spacetime metric (9.4) the only non-zero, independent Christoffel symbols are:
\begin{subequations}
\begin{eqnarray}
\Gamma_{xx}^t = \Gamma_{yy}^t = \Gamma_{zz}^t &= R\dot{R},\\
\Gamma_{tx}^x = \Gamma_{ty}^y = \Gamma_{tz}^z &= \frac{\dot{R}}{R},
\end{eqnarray}
\end{subequations}
where the dot means the time derivative. For the perfect fluid with mass density $\rho$ and pressure $p$:
\begin{equation}
T^{\mu\nu} = \rho u^\mu u^\nu + p(g^{\mu\nu} + u^\mu u^\nu),
\end{equation}
where $u^\mu$ is the four-velocity of the observer. For an observer at rest in our system of coordinates the only non-zero component of $u^\mu$ is
\begin{equation}
u^t = 1,
\end{equation}
and therefore the only non-zero components of $T^{\mu\nu}$ are:
\begin{subequations}
\begin{eqnarray}
T^{tt} &= \rho,\\
T^{xx} = T^{yy} = T^{zz} &= \frac{1}{R^2}p.
\end{eqnarray}
\end{subequations}
Eq. (9.3) may be written as;
\begin{equation}
\tilde{T}^{\mu\nu}_{\,\,\,\,,\nu} + \Gamma_{\nu\sigma}^\mu\tilde{T}^{\sigma\nu} + \Gamma_{\nu\sigma}^\nu\tilde{T}^{\mu\sigma}=0.
\end{equation}
Under the assumption that $\Lambda$, $\rho$ and $p$ all depend on the time $t$ only, it is possible to show that the  only non-trivial equation of those in Eq. (D5) is the one, where $\mu=t$. That equation takes the form:
\begin{equation}
\dot{\rho} + \frac{1}{8\pi}\dot{\Lambda} = -3\frac{\dot{R}}{R}(\rho + p).
\end{equation}
Eq. (9.5) is just the equation $\frac{1}{3}G^{tt} = \frac{8\pi}{3}\tilde{T}^{tt}$, and it is well known from the text books of general relativity. Differentiating the both sides of Eq. (9.5) with respect to the time $t$ we get:
\begin{equation}
2\frac{\dot{R}}{R}\left[\frac{\ddot{R}}{R} - \left(\frac{\dot{R}}{R}\right)^2\right] = \frac{8\pi}{3}\left(\dot{\rho} + \frac{1}{8\pi}\dot{\Lambda}\right),
\end{equation}
and Eq. (D6) implies:
\begin{equation}
\frac{\ddot{R}}{R} - \left(\frac{\dot{R}}{R}\right)^2 = -4\pi(\rho + p) = -4\pi(1+ w)\rho,
\end{equation}
where we have used Eq. (9.7) in the last equality. Solving $\rho$ from Eq. (9.5) and substituting in Eq. (D8) we get:
\begin{equation}
\frac{\ddot{R}}{R} = -\frac{1}{2}(1+3w)\left(\frac{\dot{R}}{R}\right)^2 + \frac{w+1}{2}\Lambda,
\end{equation}
which is Eq. (9.6).

\section{The Derivation of Eqs. (8.7) and (9.24)}

According to the Planck data release of the year 2015 the observed values of the Hubble constant $H_0$ and the density parameter $\Omega_m$ of the matter are: \cite{kaanee}
\begin{subequations}
\begin{eqnarray}
H_0 = (67.8\pm 0.9)kms^{-1}/MPc &=& (2.20\pm 0.03)\times 10^{-18}s^{-1},\\
\Omega_m &=& 0.308 \pm 0.012.
\end{eqnarray}
\end{subequations}
Under the assumption that the Universe is flat the density parameter associated with the dark energy is:
\begin{equation}
\Omega_\Lambda = 1 - \Omega_m = 0.692\pm 0.012.
\end{equation}
When written by means of $\Lambda$ and $H_0$ the parameter $\Omega_\Lambda$ takes the form:
\begin{equation}
\Omega_\Lambda = \frac{\Lambda}{3H_0^2},
\end{equation}
which implies:
\begin{equation}
\Lambda = 3H_0^2\Omega_\Lambda \approx 1.005\times 10^{-35}s^{-2}.
\end{equation}
The relative error in $\Lambda$ may be estimated to be:
\begin{equation}
\begin{split}
\frac{\Delta\Lambda}{\Lambda} &= \sqrt{\left(\frac{1}{\Lambda}\frac{\partial\Lambda}{\partial H_0}\right)^2(\Delta H_0)^2 + \left(\frac{1}{\Lambda}\frac{\partial\Lambda}{\partial\Omega_\Lambda}\right)^2(\Delta\Omega_\Lambda)^2}\\
 &= \sqrt{4\left(\frac{\Delta H_0}{H_0}\right)^2 + \left(\frac{\Delta\Omega_\Lambda}{\Omega_\Lambda}\right)^2}\\
&\approx 0.027,
\end{split}
\end{equation}
and so we find:
\begin{equation}
\Lambda = (1.00\pm 0.03)\times 10^{-35}s^{-2},
\end{equation}
which is Eq. (8.7).

To derive Eq. (9.24) we obtain from Eq. (9.22) an estimate for the relative error of $n$:
\begin{equation}
\begin{split}
\frac{\Delta n}{n} &= \sqrt{\left(\frac{1}{n}\frac{\partial n}{\partial H_0}\right)^2(\Delta H_0)^2 + \left(\frac{1}{n}\frac{\partial n}{\partial\Lambda}\right)^2(\Delta\Lambda)^2}\\
& = \sqrt{\left(\frac{\Lambda}{H_0^3}n\right)^2(\Delta H_0)^2 + \left(\frac{1}{2H_0^2}n\right)^2(\Delta\Lambda)^2}\\
&\approx 0.090,
\end{split}
\end{equation}
and so we find:
\begin{equation}
n = 2.1 \pm 0.2,
\end{equation}
which is Eq. (9.24).

\


\begin{thebibliography} {40}
\bibitem{kaakuu} The idea that the cosmological constant may arise from the microscopic structure of spacetime has been advocated, for example, by T. Padmanabhan and H. Padmanabhan, Int. J. Mod. Phys. {\bf D23} (2014) 1430011.

\medskip
\end{thebibliography}
\end{document}